\documentclass[conference]{IEEEtran}
\usepackage[T1]{fontenc}

\usepackage{cite}

\usepackage{amsmath,amssymb,amsfonts}
\usepackage{algorithmic}
\usepackage{graphicx}
\usepackage{textcomp}
\usepackage[dvipsnames]{xcolor}
\usepackage{xspace}
\usepackage{xargs}

\usepackage{hyperref}

\usepackage{cleveref}
\crefname{figure}{Fig.}{Figs.}
\crefname{section}{Sec.}{Secs.}

\newcommand{\etal}{\emph{et al.}\xspace}
\newcommand{\ie}{\emph{i.e.}\xspace}
\newcommand{\eg}{\emph{e.g.}\xspace}
\newcommand{\cf}{\emph{cf.}\xspace}

\newcommand{\etc}{etc.\xspace}

\newcommand{\wrt}{\emph{w.r.t.}\xspace}

\newcommand{\bgs}{BGS\xspace}
\newcommand{\CDnetPlatform}{\url{changedetection.net}\xspace}
\newcommand{\CDnetMMXII}{\emph{CDnet 2012}\xspace}
\newcommand{\CDnetMMXIV}{\emph{CDnet 2014}\xspace}

\newcommand{\numMethodsBGS}{$40$\xspace}

\newcommand{\numVideos}{$53$\xspace}

\newcommand{\first}{1\textsuperscript{st}\xspace}
\newcommand{\second}{2\textsuperscript{nd}\xspace}
\newcommand{\third}{3\textsuperscript{rd}\xspace}

\usepackage{subfig}

\begin{document}

\newcommand{\paperA}{paper~A~\cite{Pierard2024Foundations-arxiv}\xspace}
\newcommand{\paperB}{paper~B~\cite{Pierard2024TheTile-arxiv}\xspace}
\newcommand{\paperC}{paper~C~\cite{Halin2024AHitchhikers-arxiv}\xspace}
\newcommand{\PaperA}{Paper~A~\cite{Pierard2024Foundations-arxiv}\xspace}
\newcommand{\PaperB}{Paper~B~\cite{Pierard2024TheTile-arxiv}\xspace}
\newcommand{\PaperC}{Paper~C~\cite{Halin2024AHitchhikers-arxiv}\xspace}

\global\long\def\sampleSpace{\Omega}%
\global\long\def\aSample{\omega}%
\global\long\def\eventSpace{\Sigma}%
\global\long\def\anEvent{E}%
\global\long\def\measurableSpace{(\sampleSpace,\eventSpace)}%
\global\long\def\expectedValueSymbol{\mathbf{E}}%

\global\long\def\aPerformance{P}%
\global\long\def\allPerformances{\mathbb{\aPerformance}_{\measurableSpace}}%
\global\long\def\aSetOfPerformances{\Pi}%
\global\long\def\randVarSatisfaction{S}%
\global\long\def\aScore{X}%
\global\long\def\allScores{\mathbb{\aScore}_{\measurableSpace}}%
\newcommandx\domainOfScore[1][usedefault, addprefix=\global, 1=\aScore]{\mathrm{dom}(#1)}%
\global\long\def\evaluation{\mathrm{eval}}%
\global\long\def\opFilter{\mathrm{filter}_\randVarImportance}%
\global\long\def\opNoSkill{\mathrm{no\text{\textendash{}}skill}}%
\global\long\def\opPriorShift{\mathrm{shift}_{\pi\rightarrow\pi'}}%
\global\long\def\opChangePredictedClass{\mathrm{change}_{\randVarPredictedClass}}%
\global\long\def\opChangeGroundtruthClass{\mathrm{change}_{\randVarGroundtruthClass}}%
\global\long\def\opSwapGroundtruthAndPredictedClasses{\mathrm{swap}_{\randVarGroundtruthClass\leftrightarrow\randVarPredictedClass}}%
\global\long\def\opSwapClasses{\mathrm{swap}_{\classNeg\leftrightarrow\classPos}}%
\global\long\def\allWorstPerformances{\frownie}
\global\long\def\allBestPerformances{\smiley}

\global\long\def\randVarGroundtruthClass{Y}%
\global\long\def\randVarPredictedClass{\hat{Y}}%
\global\long\def\allClasses{\mathbb{C}}%
\global\long\def\aClass{c}%
\global\long\def\classNeg{c_-}%
\global\long\def\classPos{c_+}%
\global\long\def\sampleTN{tn}%
\global\long\def\sampleFP{fp}%
\global\long\def\sampleFN{fn}%
\global\long\def\sampleTP{tp}%
\global\long\def\eventTN{\{\sampleTN\}}%
\global\long\def\eventFP{\{\sampleFP\}}%
\global\long\def\eventFN{\{\sampleFN\}}%
\global\long\def\eventTP{\{\sampleTP\}}%
\global\long\def\scorePTN{PTN}%
\global\long\def\scorePFP{PFP}%
\global\long\def\scorePFN{PFN}%
\global\long\def\scorePTP{PTP}%
\global\long\def\scoreAccuracy{A}%
\global\long\def\scoreExpectedSatisfaction{\aScore_{\randVarSatisfaction}}%
\global\long\def\scoreTNR{TNR}%
\global\long\def\scoreFPR{FPR}%
\global\long\def\scoreTPR{TPR}%
\global\long\def\scoreFNR{FNR}%
\global\long\def\scoreNPV{NPV}%
\global\long\def\scoreFOR{FOR}%
\global\long\def\scorePPV{PPV}%
\global\long\def\scorePrecision{\scorePPV}%
\global\long\def\scoreFDR{FDR}%
\global\long\def\scoreJaccardNeg{J_-}%
\global\long\def\scoreJaccardPos{J_+}%
\global\long\def\scoreCohenKappa{\kappa}%
\global\long\def\scoreScottPi{\pi}%
\global\long\def\scoreFleissKappa{\kappa}%
\global\long\def\scoreBalancedAccuracy{BA}%
\global\long\def\scoreWeightedAccuracy{WA}%
\global\long\def\scoreYoudenJ{J_Y}
\global\long\def\scorePLR{PLR}%
\global\long\def\scoreNLR{NLR}%
\global\long\def\scoreOR{OR}%
\global\long\def\scoreSNPV{SNPV}%
\global\long\def\scoreSPPV{SPPV}%
\global\long\def\scoreACP{ACP}%
\global\long\def\scoreFOne{F_{1}}%
\newcommandx\scoreFBeta[1][usedefault, addprefix=\global, 1=\beta]{F_{#1}}%
\global\long\def\scoreFOne{\scoreFBeta[1]}%
\global\long\def\priorpos{\pi_+}%
\global\long\def\priorneg{\pi_-}%
\global\long\def\scoreBiasIndex{BI}%
\global\long\def\ratepos{\tau_+}%
\global\long\def\rateneg{\tau_-}%
\global\long\def\scoreACP{ACP}%
\global\long\def\scorePFour{P_4}%
\global\long\def\normalizedConfusionMatrix{C}%
\global\long\def\scoreConfusionMatrixDeterminant{|\normalizedConfusionMatrix|}

\global\long\def\allEntities{\mathbb{E}}%
\global\long\def\entitiesToRank{\mathbb{E}}%
\global\long\def\anEntity{\epsilon}%
\global\long\def\randVarImportance{I}%
\global\long\def\randVarCanonicalImportance{\randVarImportance_{a,b}}
\global\long\def\canonicalRankingScore{\rankingScore[\randVarCanonicalImportance]}
\newcommandx\rankingScore[1][usedefault, addprefix=\global, 1=\randVarImportance]{R_{#1}}%
\global\long\def\canonicalRankingScore{\rankingScore[\randVarImportance_{a,b}]}
\global\long\def\scoreVUT{VUT}%
\global\long\def\tileCurvePriors{\gamma_\pi}
\global\long\def\tileCurveRates{\gamma_\tau}
\global\long\def\relWorseOrEquivalent{\lesssim}%
\global\long\def\relBetterOrEquivalent{\gtrsim}%
\global\long\def\relEquivalent{\sim}%
\global\long\def\relBetter{>}%
\global\long\def\relWorse{<}%
\global\long\def\relIncomparable{\not\lesseqqgtr}%
\global\long\def\rank{\mathrm{rank}_\entitiesToRank}%
\global\long\def\ordering{\relWorseOrEquivalent}%
\global\long\def\invertedOrdering{\relBetterOrEquivalent}%

\global\long\def\LScityscapes{\ding{171}}
\global\long\def\LSade{\ding{170}}
\global\long\def\LSvoc{\ding{169}}
\global\long\def\LScoco{\ding{168}}

\global\long\def\indicatorSymbol{\mathbf{1}}
\global\long\def\realNumbers{\mathbb{R}}%
\global\long\def\aRelation{\mathcal{R}}%
\global\long\def\achievableByCombinations{\Phi}%
\global\long\def\allConvexCombinations{\mathrm{conv}}%
\newcommand{\indep}{\perp \!\!\! \perp}


\global\long\def\cityscapes{\LScityscapes{}~Cityscapes}
\global\long\def\ade{\LSade{}~ADE20K}
\global\long\def\voc{\LSvoc{}~Pascal VOC 2012}
\global\long\def\coco{\LScoco{}~COCO-Stuff 164k}

\newcommand{\MethodDesigner}{Bernadette}
\newcommand{\Benchmarker}{Leonard}
\newcommand{\AppDeveloper}{Howard}
\newcommand{\TheoreticalAnalyst}{Sheldon}

\newcommand{\tile}{Tile\xspace}
\newcommand{\tiles}{Tiles\xspace}
\newcommand{\valueTile}{Value Tile\xspace}
\newcommand{\valueTiles}{Value Tiles\xspace}
\newcommand{\baselineTile}{Baseline Value Tile\xspace}
\newcommand{\baselineTiles}{Baseline Value Tiles\xspace}
\newcommand{\SOTATile}{State-of-the-Art Value Tile\xspace}
\newcommand{\SOTATiles}{State-of-the-Art Value Tiles\xspace}
\newcommand{\noSkillTile}{No-Skill Tile\xspace}
\newcommand{\noSkillTiles}{No-Skill Tiles\xspace}
\newcommand{\skillTile}{Relative-Skill Tile\xspace}
\newcommand{\skillTiles}{Relative-Skill Tiles\xspace}
\newcommand{\correlationTile}{Correlation Tile\xspace}
\newcommand{\correlationTiles}{Correlation Tiles\xspace}
\newcommand{\rankingTile}{Ranking Tile\xspace}
\newcommand{\rankingTiles}{Ranking Tiles\xspace}
\newcommand{\entityTile}{Entity Tile\xspace}
\newcommand{\entityTiles}{Entity Tiles\xspace}

\global\long\def\aNonSkilledPerformance{\aPerformance_{\indep}}
\global\long\def\allNonSkilledPerformances{\mathbb{\aPerformance}^{\randVarGroundtruthClass\indep\randVarPredictedClass}_{\measurableSpace}}%

\global\long\def\allPriorFixedPerformances{\mathbb{\aPerformance}^{\priorpos}_{\measurableSpace}}%

\newcommand{\comma}{\,,}
\newcommand{\point}{\,.}

\newcommandx\unconditionalProbabilisticScore[1]{\aScore_{#1}^{U}}%
\global\long\def\formulaPTN{\unconditionalProbabilisticScore{\eventTN}}%
\global\long\def\formulaPFP{\unconditionalProbabilisticScore{\eventFP}}%
\global\long\def\formulaPFN{\unconditionalProbabilisticScore{\eventFN}}%
\global\long\def\formulaPTP{\unconditionalProbabilisticScore{\eventTP}}%
\global\long\def\formulapriorneg{\unconditionalProbabilisticScore{\{\sampleTN,\sampleFP\}}}%
\global\long\def\formulapriorpos{\unconditionalProbabilisticScore{\{\sampleFN,\sampleTP\}}}%
\global\long\def\formularateneg{\unconditionalProbabilisticScore{\{\sampleTN,\sampleFN\}}}%
\global\long\def\formularatepos{\unconditionalProbabilisticScore{\{\sampleFP,\sampleTP\}}}%
\global\long\def\formulaAccuracy{\unconditionalProbabilisticScore{\{\sampleTN,\sampleTP\}}}%

\newcommandx\conditionalProbabilisticScore[2]{\aScore_{#1 \vert #2}^{C}}%
\global\long\def\formulaTNR{\conditionalProbabilisticScore{\{\sampleTN\}}{\{\sampleTN,\sampleFP\}}}%
\global\long\def\formulaTPR{\conditionalProbabilisticScore{\{\sampleTP\}}{\{\sampleFN,\sampleTP\}}}%
\global\long\def\formulaNPV{\conditionalProbabilisticScore{\{\sampleTN\}}{\{\sampleTN,\sampleFN\}}}%
\global\long\def\formulaPPV{\conditionalProbabilisticScore{\{\sampleTP\}}{\{\sampleFP,\sampleTP\}}}%
\global\long\def\formulaJaccardNeg{\conditionalProbabilisticScore{\{\sampleTN\}}{\{\sampleTN,\sampleFP,\sampleFN\}}}%
\global\long\def\formulaJaccardPos{\conditionalProbabilisticScore{\{\sampleTP\}}{\{\sampleFP,\sampleFN,\sampleTP\}}}%

\global\long\def\scoreBennettS{S}


\title{A Methodology to Evaluate Strategies Predicting Rankings on Unseen Domains}

\author{S\'ebastien Pi\'erard, Adrien Deli\`ege, Ana\"is Halin, Marc Van Droogenbroeck\\
Montefiore Institute, University of Li\`ege, Belgium\\
{\tt\small \{S.Pierard,Adrien.Deliege,Anais.Halin,M.VanDroogenbroeck\}@uliege.be}
}

\maketitle


\begin{abstract}

Frequently, multiple entities (methods, algorithms, procedures, solutions, \etc) can be developed for a common task and applied across various domains that differ in the distribution of scenarios encountered. For example, in computer vision, the input data provided to image analysis methods depend on the type of sensor used, its location, and the scene content. However, a crucial difficulty remains: can we predict which entities will perform best in a new domain based on assessments on known domains, without having to carry out new and costly evaluations? This paper presents an original methodology to address this question, in a leave-one-domain-out fashion, for various application-specific preferences. We illustrate its use with $30$ strategies to predict the rankings of \numMethodsBGS entities (unsupervised background subtraction methods) on \numVideos domains (videos).

\end{abstract}


\section{Introduction}
\label{sec:intro}

\begin{figure}
    \centering
    \includegraphics[width=0.9\linewidth]{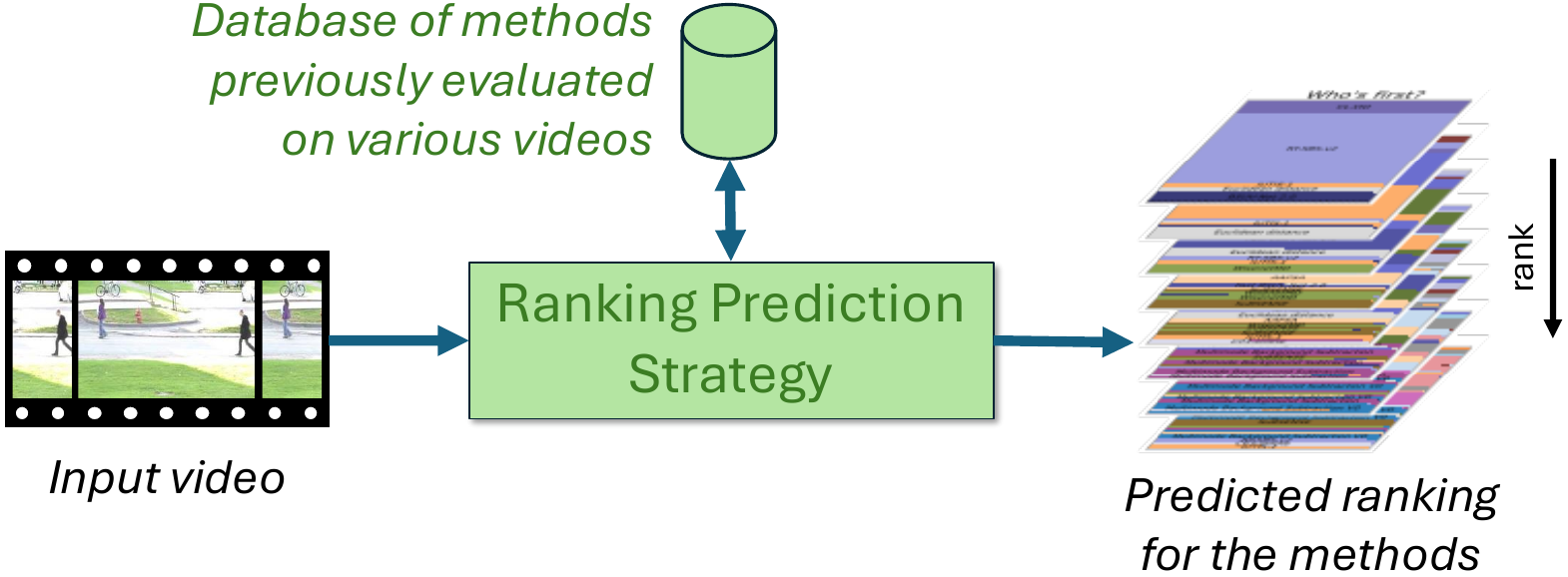}
    \caption{In this paper, we explore the problem of predicting the rankings of computer vision methods (we take the particular case of background subtraction methods) on any new domain (video) based on a database storing the performances of these methods, previously evaluated on other domains (videos).}
    \label{fig:graphical-abstract}
\end{figure}

Why do computer scientists rank entities (methods, algorithms, \etc)? Certainly, not just for picking a winner for a contest. The aim is rather to gain some knowledge that can help to select the most promising entity for a particular application. Predicting a ranking is a topic that has rarely been studied. As depicted in \cref{fig:graphical-abstract}, we explore the case of a computer vision task and present an original methodology for investigating this theme, based on recent theory \cite{Pierard2025Foundations} and tools \cite{Pierard2024TheTile-arxiv,Halin2024AHitchhikers-arxiv} that are briefly recalled.

Background subtraction (\bgs) is a computer-vision task for which the established rankings drastically differ from a video to another. Undoubtedly, this is due, on the one hand, to the variety of principles supporting these methods and, on the other hand, to the wide range of potential applications, and so to the wide variety of videos on which these methods can be applied. As a result, practitioners are frequently confronted with the selection of the most appropriate method for new video sequences, new camera placements, or video streams with characteristics changing unpredictably over time.

A first path consists in predicting the same ranking for all applicative domains, regardless of their characteristics. This is expected to be suboptimal, but it is worth trying such strategies for the sake of simplicity.

A second path consists in predicting the rankings based on the knowledge the practitioner has about his/her applicative domain. For example, he/she can specify a category to which all the inputs belong. The dataset \CDnetMMXII~\cite{Goyette2012Changedetection} defined $6$ categories: \emph{baseline}, \emph{dynamic background}, \emph{camera jitter}, \emph{intermittent object motion}, \emph{shadow}, and \emph{thermal}, later complemented in \CDnetMMXIV~\cite{Wang2014CDnet} by \emph{bad weather}, \emph{low framerate}, \emph{night videos}, \emph{PTZ}, and \emph{turbulence}.

A third path consists in analyzing the input and predicting the ranking based on the perceived domain characteristics. Practitioners can collect representative images and analyze them to characterize their applicative domains as statistical distributions. For example, Halin \etal~\cite{Halin2025Physically} have demonstrated the feasibility of characterizing domains, based on images, in a physically interpretable way with simulation-based inference~\cite{Cranmer2020TheFrontier} techniques. Alternatively, one could also classify pixels (\eg, with semantic segmentation) and characterize the domain by the predicted distribution of classes.

In fact, our \bgs example is just a particular case of a general and common problem that consists in predicting the rankings of entities for unseen domains based on performance evaluations previously performed for other domains, without having to carry out any new and costly evaluation. We go beyond just predicting the ``best'' entity as a user could be constrained to any subset of entities. To evaluate a predictive strategy, we compute the probability that, for any randomly chosen pair of entities $(\anEntity_a,\anEntity_b)$, $\anEntity_a$ is predicted as worse or better than $\anEntity_b$ when it is actually the case.

Currently, scientists are not aware of how well simple and intuitive strategies perform. The ability to exploit existing rankings is all a blur. Our objective is to provide methodological elements and rigorous tools to clarify this. An originality of this work is that we do not rely on a small set of scores, but on a large family of scores, all of which have the required properties to induce meaningful rankings covering a diversity of application-specific preferences.

Our two main contributions are the following. 
(1) First, we establish a new and original methodology to evaluate and compare strategies predicting rankings on unseen domains. It leverages the recently introduced theoretical foundations for performance-based ranking~\cite{Pierard2025Foundations} as well as a recently described tool (\ie, the \tile~\cite{Pierard2024TheTile-arxiv,Halin2024AHitchhikers-arxiv}). This framework allows us to cover an infinite and diversified family of meaningful performance-based rankings and to see how strategies perform according to application-specific preferences. 
(2) Second, we show a new usage for the \tile, originally introduced as a graphical tool to analyze and compare performances of two-class classifiers. In this work, the \tile appears to be the appropriate tool to depict the performance of strategies predicting rankings of two-class classifiers on unseen domains, and to compare them.

\section{Related Work}

\subsection{Background Subtraction}

Most of the literature on \bgs is about the  methods \cite{Bouwmans2019Deep,Bouwmans2019VisualHuman,Bouwmans2019VisualNatural}, with some intended to be all-purpose (\ie, universal) while others are developed for specific application domains~\cite{Braham2016Deep}. Some consider constraints such as limited resources (time, memory, \etc) \cite{Barnich2011ViBe,Cioppa2020RealTime}, and some tackle multimodality \cite{Leens2009Combining}.

A small part of the literature also concerns datasets, initially used to compare the \bgs methods, then leveraged to develop supervised methods~\cite{Bouwmans2019Deep}. The best-known datasets are \CDnetMMXII~\cite{Goyette2012Changedetection} and its extension \CDnetMMXIV~\cite{Wang2014CDnet}, which are integrated into the \CDnetPlatform platform.

Eventually, a part of the literature focuses on performance analysis. \bgs is typically identified to a pixel-based two-class classification, where the negative and positive classes correspond to the background and the foreground (in the following, we denote a true negative, false positive, false negative, and true positive by $\sampleTN$, $\sampleFP$, $\sampleFN$, and $\sampleTP$). 
Jodoin \etal~\cite{Jodoin2014Overview} and Piérard \etal~\cite{Pierar2022AnExploration-arxiv} determined which performances are achievable by combining existing \bgs methods. Braham \etal~\cite{Braham2017Semantic} introduced a generic technique to improve \bgs methods performances. Piérard \etal~\cite{Pierard2015APerfect} studied the limits of achievable performances with simple methods (pixelwise color comparisons), and how they vary with some properties (\eg, amounts of noise and shadows) of the videos analyzed. Piérard \etal~\cite{Pierard2020Summarizing} highlighted the pitfalls of averaging values of scores measured on several videos, and provided a technique for summarizing the corresponding performances into a single interpretable one.

\subsection{Performance-Based Ranking}

Theoretical foundations for performance-based rankings, grounded in probability and order theories, have recently been introduced by Piérard \etal~\cite{Pierard2025Foundations} through a rigorous axiomatic framework\footnote{
    Their \first axiom states that any performance-based ranking should be derived from a preorder on performances, which ensures the stability of the rankings \wrt insertions and deletions of ranked entities. Their \second axiom gives compatibility conditions between the preorders and the considered task, modeled by a random variable called \emph{satisfaction}. Their \third axiom gives compatibility conditions between the preorders and known properties about the evaluation (\ie, the mapping of the entities to their performances).
}. These axioms are guardrails to guarantee meaningful rankings while leaving the flexibility to adjust the rankings \wrt application-specific preferences.

\begin{figure}
\begin{centering}
\includegraphics[width=\linewidth]{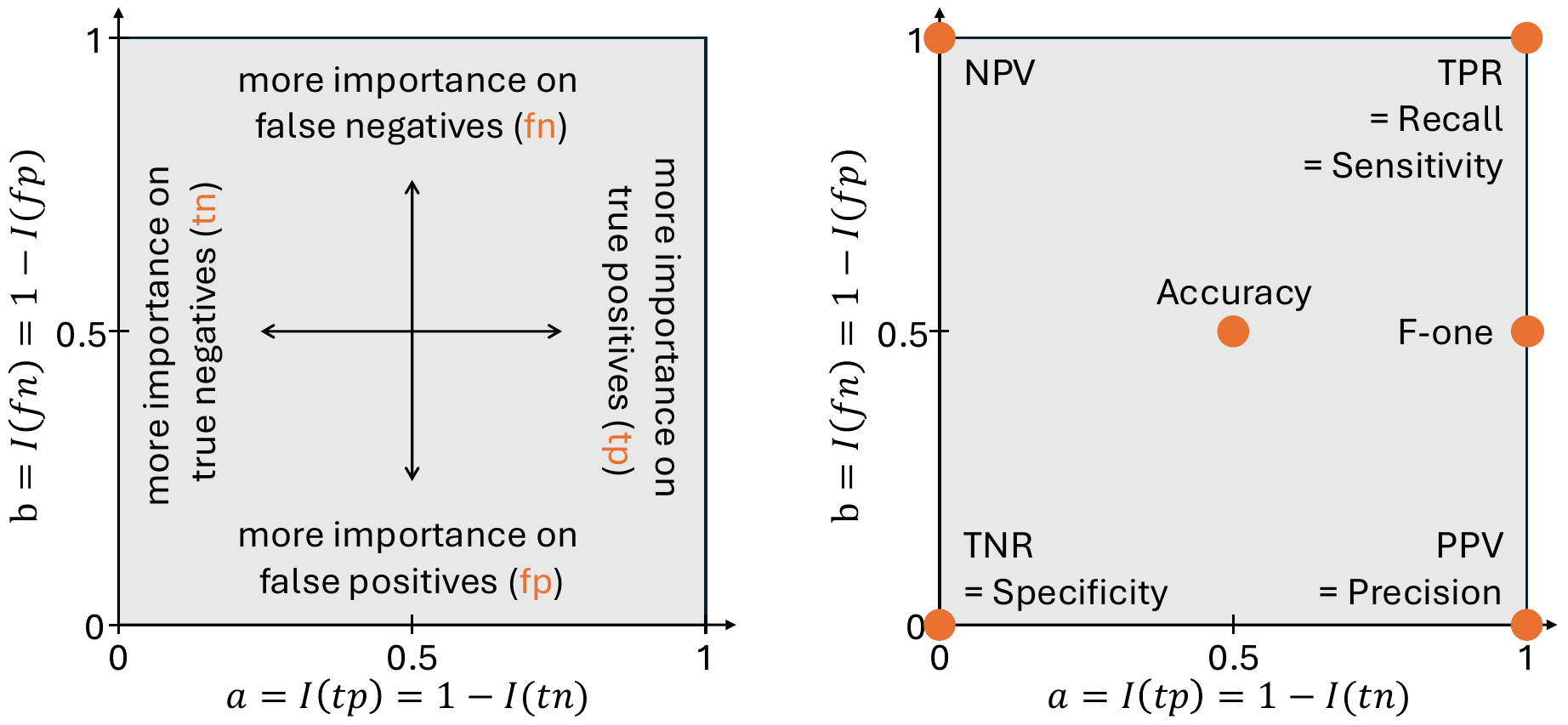}
\par\end{centering}
\caption{
    Two equivalent readings of the \tile: a map of application-specific importances (left) and a map of scores to induce meaningful performance-based rankings (right).
    \label{fig:tile}
}
\end{figure}

When particularized to two-class classification (\eg, between background and foreground, or between negatives and positives), it has been shown in~\cite{Pierard2025Foundations,Pierard2024TheTile-arxiv,Halin2024AHitchhikers-arxiv} that one can fine-tune the relative importance given to the two types of correct classifications, as well as to the two types of erroneous classifications, by inducing the performance ordering from a parametric score $\rankingScore$ called \emph{ranking score}:
\begin{equation}
    \rankingScore:\aPerformance\mapsto\rankingScore(\aPerformance)=\frac{
        \sum_{\aSample\in\{\sampleTN,\sampleTP\}} 
        \randVarImportance(\aSample) \aPerformance(\{\aSample\})
    }{
        \sum_{\aSample\in\{\sampleTN,\sampleFP,\sampleFN,\sampleTP\}} 
        \randVarImportance(\aSample) \aPerformance(\{\aSample\})
    } \comma
    \label{eq:ranking-scores}
\end{equation}
where $\aPerformance$ denotes the performance, and $\randVarImportance$ the importance, \ie, the application-specific preferences. Ranking scores sharing the same values for $a=\frac{
    \randVarImportance(\sampleTP)
}{
    \randVarImportance(\sampleTN)
    +\randVarImportance(\sampleTP)
}$ and $b=\frac{
    \randVarImportance(\sampleFN)
}{
    \randVarImportance(\sampleFP)
    +\randVarImportance(\sampleFN)
}$ lead to the same rankings~\cite{Pierard2024TheTile-arxiv}. So, the diversity of rankings can be shown on the $(a,b)\in[0,1]^2$ square, coined as the \tile~\cite{Pierard2024TheTile-arxiv,Halin2024AHitchhikers-arxiv}. Interestingly, the family of ranking scores contains scores commonly used in the \bgs community (see \cref{fig:tile}).

\subsection{Ranking Prediction}

Predicting a ranking of entities, sometimes called ``learning to rank'' in machine learning, is a task used for instance in recommendation systems, or action quality assessment (\eg, in sports). 
The strategies generally fall into three categories: \textit{pointwise}, where strategies predict the performance score of each entity individually, and derive a ranking from these scores~\cite{Cossock2006Subset,Li2007McRank}; \textit{pairwise}, where strategies predict the relative ordering between entity pairs, and derive a ranking from pairwise orderings~\cite{Burges2005LearningToRank}; and \textit{listwise}, where strategies predict directly the entire ranked list~\cite{Cao2007LearningToRank}. Pointwise approaches are simple and efficient but ignore the relative nature of ranking tasks. Pairwise strategies better capture it and are widely used, though they struggle with global ranking consistency. Listwise approaches, by directly optimizing ranking scores, often achieve superior performance, but are typically more complex and computationally demanding~\cite{Liu2009Learning,Tax2015ACrossbenchmark}. Compared to that literature, we do not aim to learn a model that ranks entities based on domain (videos) features; instead, we leverage pre-computed rankings across multiple domains to infer the ranking on a previously unseen one.

\begin{figure*}[t!]
\global\long\def\strategyCDnet{\ensuremath{\mathrm{CDnet}}}
\global\long\def\strategySemanticDistance{\ensuremath{\mathrm{sem-d}^\dagger}}
\global\long\def\strategySemanticDistanceCategory{\ensuremath{\mathrm{sem-d}^{\dagger*}}}
\begin{centering}
    \hfill{}
    \subfloat[Strategy \strategyCDnet]{
        \begin{centering}
        \includegraphics[width=0.4\columnwidth]{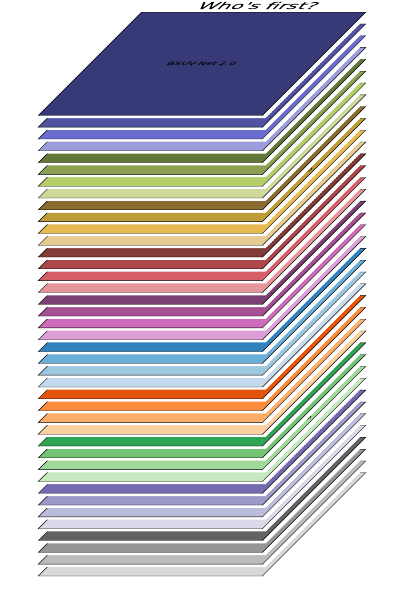}
        \par\end{centering}
    }
    \hfill{}\hfill{}
    \subfloat[Strategy \strategySemanticDistance]{
        \begin{centering}
        \includegraphics[width=0.4\columnwidth]{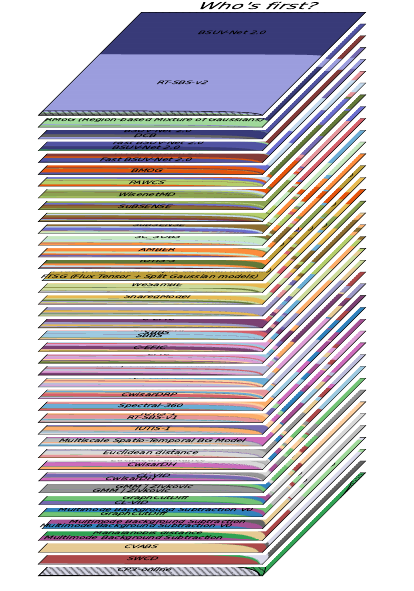}
        \par\end{centering}
    }
    \hfill{}\hfill{}
    \subfloat[Strategy \strategySemanticDistanceCategory]{
        \begin{centering}
        \includegraphics[width=0.4\columnwidth]{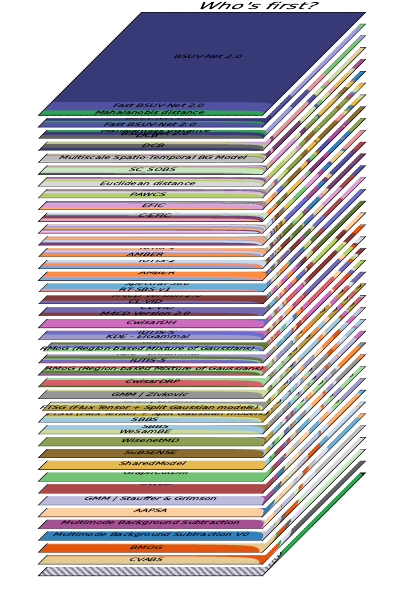}
        \par\end{centering}
    }
    \hfill{}\hfill{}
    \subfloat[Real rankings]{
        \begin{centering}
        \includegraphics[width=0.4\columnwidth]{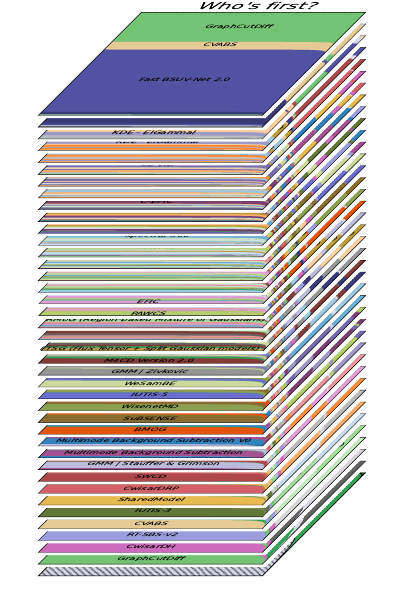}
        \par\end{centering}
    }
    \hfill{}
\par\end{centering}
\begin{centering}
    \hfill{}
    \subfloat[Strategy \strategyCDnet]{
        \begin{centering}
        \begin{minipage}[t]{0.39\columnwidth}%
            \begin{center}
            \includegraphics[scale=0.27]{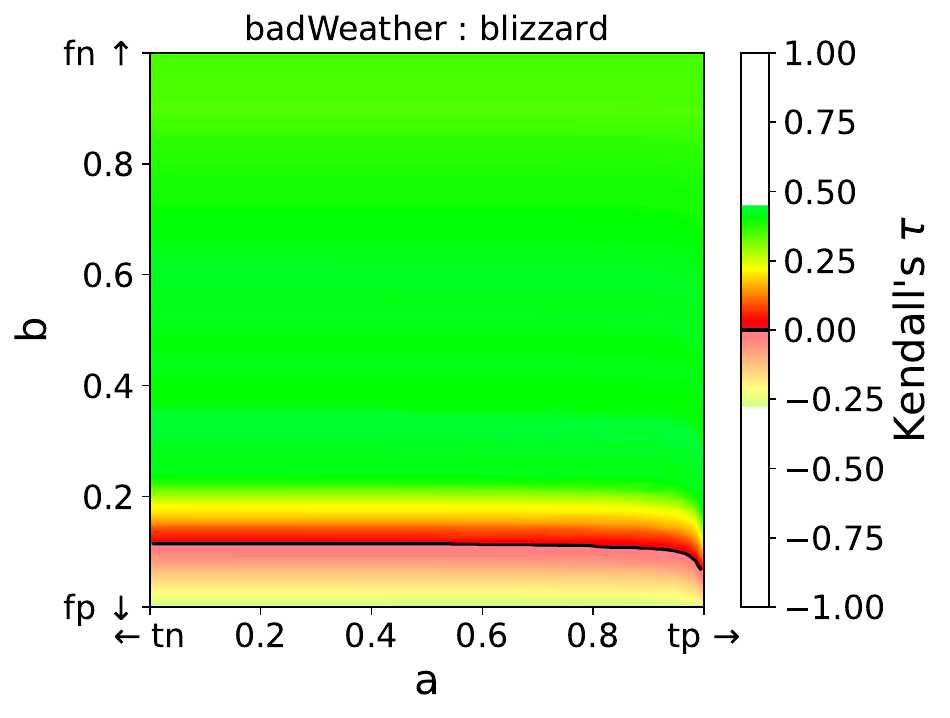}
            \par\end{center}%
        \end{minipage}
        \par\end{centering}
    }
    \hfill{}\hfill{}
    \subfloat[Strategy \strategySemanticDistance]{
        \begin{centering}
        \begin{minipage}[t]{0.39\columnwidth}%
            \begin{center}
            \includegraphics[scale=0.27]{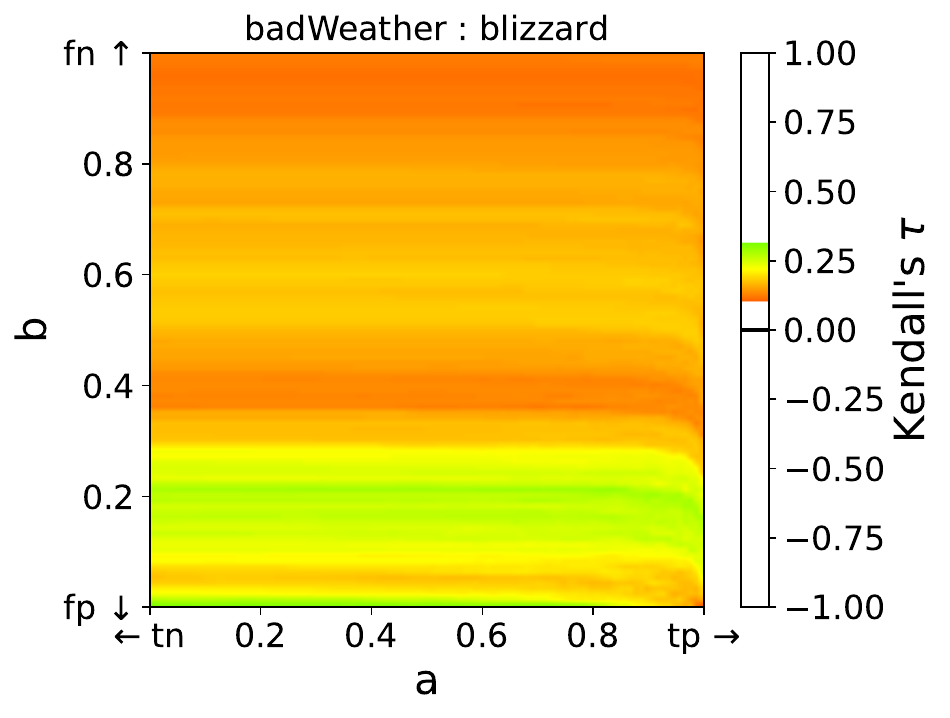}
            \par\end{center}%
        \end{minipage}
        \par\end{centering}
    }
    \hfill{}\hfill{}
    \subfloat[Strategy \strategySemanticDistanceCategory]{
        \begin{centering}
        \begin{minipage}[t]{0.39\columnwidth}%
            \begin{center}
            \includegraphics[scale=0.27]{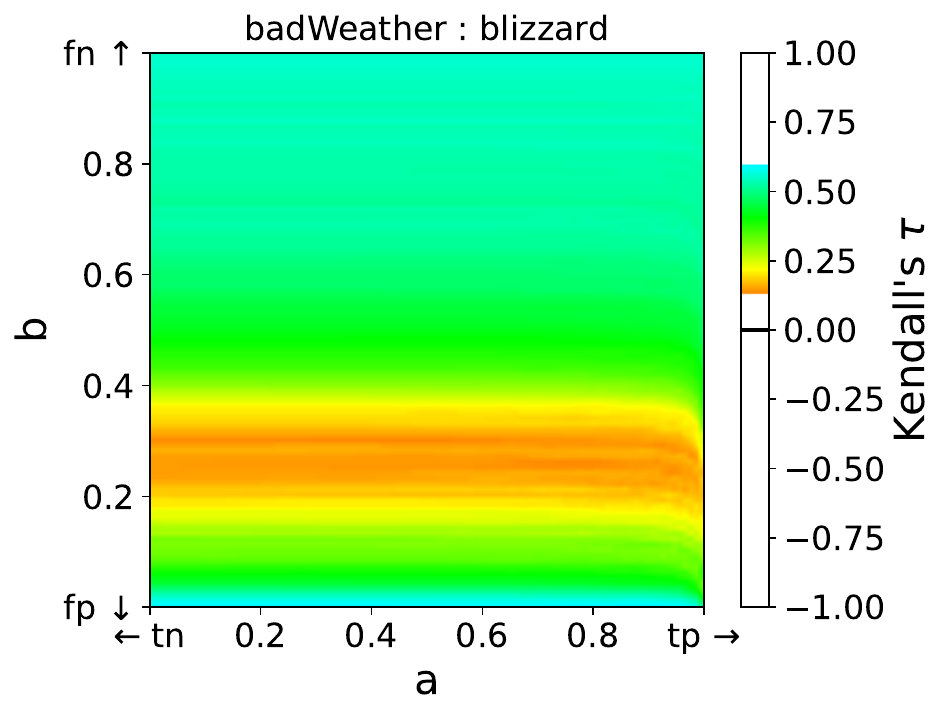}
            \par\end{center}%
        \end{minipage}
        \par\end{centering}
    }
    \hfill{}\hfill{}
    \subfloat[Choice \wrt $\randVarImportance$]{
        \begin{centering}
        \begin{minipage}[t]{0.39\columnwidth}%
            \begin{center}
            \includegraphics[scale=0.27]{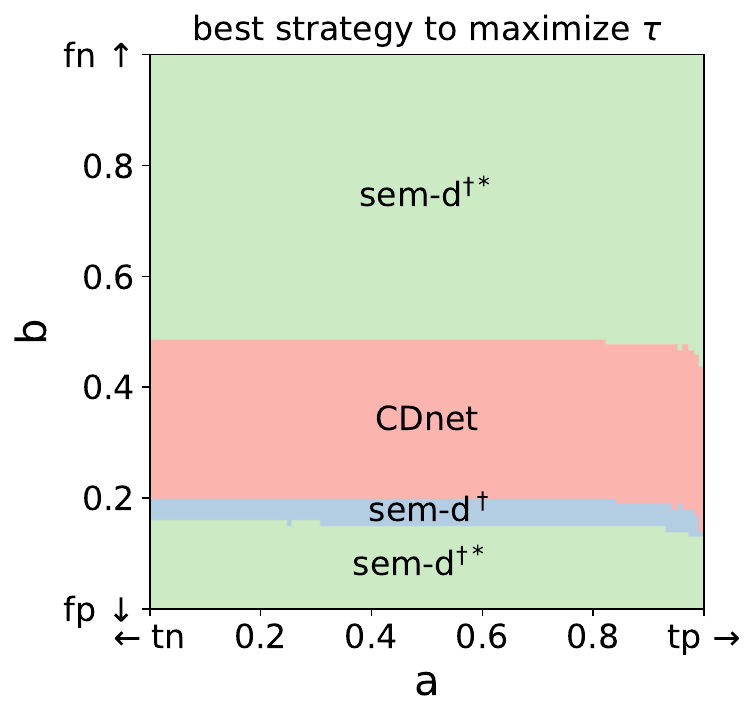}
            \par\end{center}%
        \end{minipage}
        \par\end{centering}
    }
    \hfill{}
\par\end{centering}
\caption{
    Illustration of our methodology for the evaluation and comparison of $3$ strategies to predict the rankings of $40$ \bgs methods on the video ``bad weather: blizzard''. We use \tiles~\cite{Pierard2024TheTile-arxiv} to cover the different application-specific preferences $(a,b)$. The upper row shows: (a)~the predicted ranking based just on the global ranking given on \CDnetPlatform  (strategy \strategyCDnet), (b)~the one based on the performance measured on another semantically close video (strategy \strategySemanticDistance), (c)~the one based on the performance measured on another semantically close video in the same category (strategy \strategySemanticDistanceCategory), and (d)~the actual ranking we would like to predict (our ground truth). These ``mille-feuilles'' are stackings of entity \tiles \cite{Halin2024AHitchhikers-arxiv}: the k-th layer shows the methods ranked k-th, the worst methods being at the base of the mille-feuille, and the best ones on its top. The lower row shows the correlation \tiles \cite{Halin2024AHitchhikers-arxiv} between these rankings: (e)~the correlation between (a) and (d), (f)~the correlation between (b) and (d), and (g)~ the correlation between (c) and (d). The \tile (h) shows which strategy gives the best correlation. This methodology is applied in \cref{sec:application} to compare many more strategies on a diversified set of \numVideos videos.
    \label{fig:experiment_introductory_example_blizzard_S}
}
\end{figure*}
\section{Methodology}
\label{sec:methodology}

We now describe our new methodology to evaluate and compare strategies predicting the rankings of methods in unseen domains. We focus on strategies able to predict a large family of meaningful rankings: all those induced by ranking scores~\cite{Pierard2025Foundations} (\cf Eq.~\ref{eq:ranking-scores}). For problems comparable to  two-class classification, these rankings can be mapped on the  \tile~\cite{Pierard2024TheTile-arxiv,Halin2024AHitchhikers-arxiv}, which leads to a \emph{mille-feuille}\footnote{\emph{Mille-feuille} pastries are composed of several layers, with some made with 
puff pastry, resembling to a stacking of multiple sheets.}, as shown in \cref{fig:experiment_introductory_example_blizzard_S}.

\begin{figure}
\center 
\includegraphics[width=0.7\linewidth]{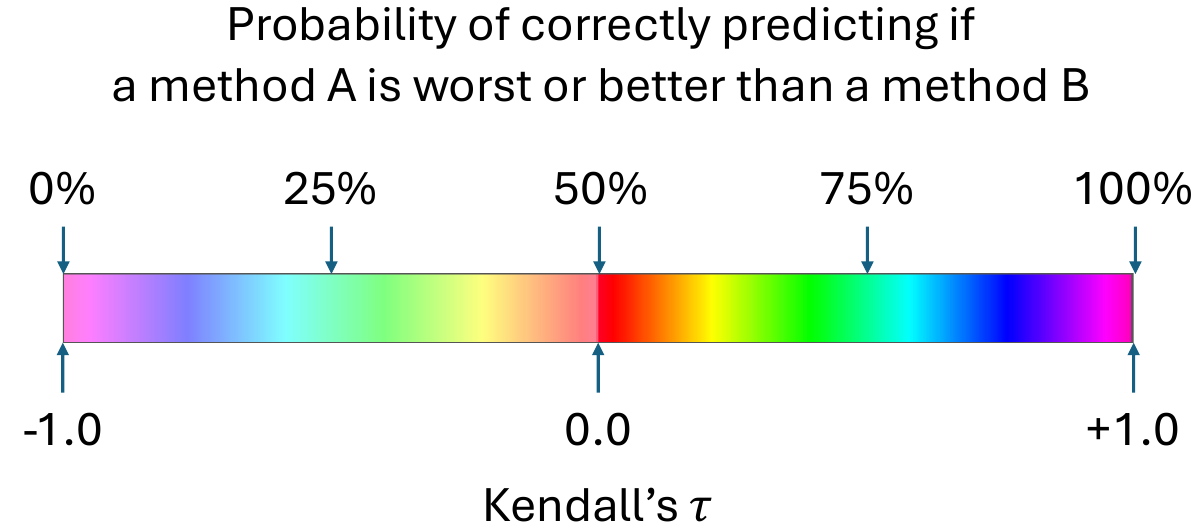}

\caption{Interpretation of the rank correlation $\tau$ (Kendall). 
Note the color code for displaying the value of $\tau$, as is used in other figures.
\label{fig:colormap-for-correlations}}

\end{figure}

We propose to evaluate strategies, for any given domain, with correlation \tiles \cite{Halin2024AHitchhikers-arxiv} giving the rank-correlation between the predicted and ground-truth rankings, \wrt the various application-specific preferences (specified by $\randVarImportance$, or $a$ and $b$). We choose Kendall's $\tau$ as this score is coherent with the performance-based ranking framework\footnote{
    It turns out that Kendall's $\tau$ 
    is itself a ranking score for ranking problems when pairs of miss-ordered and well-ordered methods are assigned a satisfaction of -1 and +1, respectively. See the appendix of~\cite{Pierard2025Foundations} for more information.
}. Moreover, Kendall's $\tau$ can be nicely interpreted as the probability of observing, in two given rankings, the same relative order between two randomly chosen methods, this probability being linearly mapped into the $[-1,+1]$ interval (see \cref{fig:colormap-for-correlations}). 

Considering several domains can be done in a \emph{leave-one-domain-out} fashion, predicting the ranking for each domain using information from the others. The expected performance of the strategy is given by the \emph{mean correlation \tile}, whereas the worst-case is given by the \emph{minimum correlation \tile}. This will be illustrated in Sec.~\ref{sec:application}.

Two natural baselines can be defined when the ground-truth rankings are known for several domains. (1) The mean of $\tau$ over all domain pairs estimates the expected agreement when generalizing a ranking from a randomly chosen reference domain to another random domain. (2) The minimum of $\tau$ across all domain pairs captures the worst-case agreement for any fixed domain and arbitrary reference.

To compare various strategies, it is essential to bear in mind that the best strategy might depend on the application-specific preferences $\randVarImportance$. This can be mapped on the \tile.

\section{Application}
\label{sec:application}

We now apply our methodology to the problem of predicting the performance-based rankings of \numMethodsBGS unsupervised \bgs methods on the \numVideos videos of \CDnetMMXIV. As said, we use only rankings, no implementation of these methods, ground-truth segmentation masks, nor any inference and evaluation step. In particular, the priors and the rates of predictions for the background and foreground are unknown. We compare strategies that have no knowledge about the input video (their expected distribution put aside) and strategies that leverage some domain characterization: a category (denoted by~$^*$), a distribution of semantic classes\footnote{
    The semantic labels are used only to cluster the pixels. Thus, we can rely on predicted (not perfect) segmentations. We selected a model from MMSegmentation~\cite{Mmsegmentation2020} based on two criteria: semantic diversity and expected performance. For the former, we limited the choice to ADE20K~\cite{Zhou2017Scene} models and COCO-Stuff~\cite{Caesar2018COCO} models. Based on performance insights from \emph {entity \tiles}~\cite{Halin2024AHitchhikers-arxiv}, we selected Mask2Former~\cite{Cheng2022Masked} trained on ADE20K. 
} (denoted by~$^\dagger$), or both.

The information known about the test video (\ie, for which we must predict the ranking) is its two-fold domain characterization: (1) its category in \CDnetMMXIV and (2) an estimate of the statistical distribution of the pixel-based semantic labels. The information known for the others is three-fold: (1) its category, (2) the join distribution of the pixel-based semantic labels, ground-truth class, and estimated classes, and (3) the performance-based ranking of the \numMethodsBGS \bgs methods. All distributions are for the publicly available subset of pixels annotated as background or foreground in \CDnetMMXIV.

\subsection{Baselines for Ranking Prediction Strategies}

\begin{figure}
\center 
\includegraphics[scale=0.24]{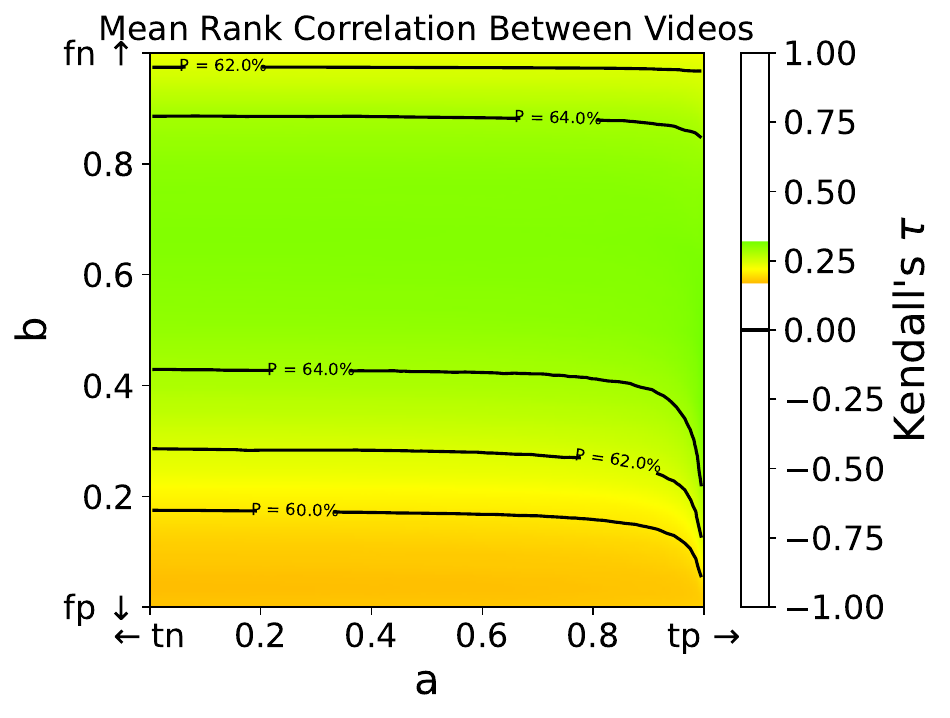}
\includegraphics[scale=0.24]{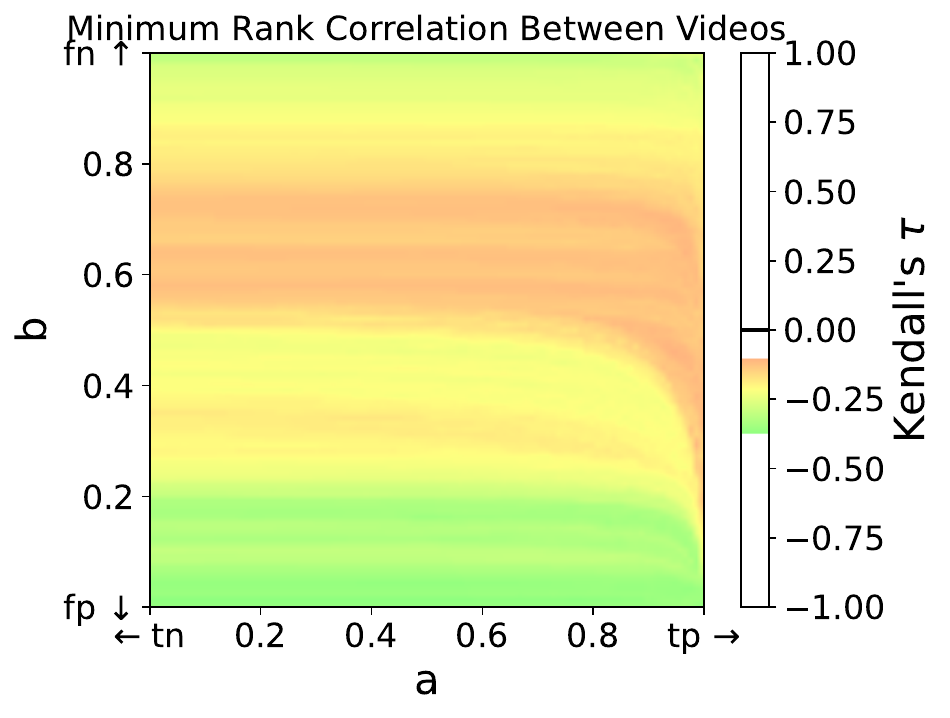}

\caption{The two natural baselines for 
ranking prediction strategies (as defined in our methodology) in the particular case of \numMethodsBGS unsupervised \bgs algorithms ranked on \numVideos videos.\label{fig:baseline}}

\end{figure}

The two natural baselines defined in  \cref{sec:methodology} are shown in  \cref{fig:baseline}. On the left-hand side, we see that there is only a little average correlation between rankings from one video to another. 
The mean value for $\tau$ around $0.25$ means that there is about $62.5\%$ chance of correctly ordering two methods if we refer to a predetermined ranking for any randomly chosen video. 
Moreover, we see that the difficulty of predicting rankings is not uniform: it seems to be a more difficult problem when $\randVarImportance(\sampleFN)$ is low \wrt $\randVarImportance(\sampleFP)$ (bottom of the \tile). 
On the right-hand side, we observe that, in the worst case, regardless of what the application-specific preferences $\randVarImportance$ are, there can be a negative correlation between the rankings on two videos. All things considered, the problem of predicting the performance-based rankings of \bgs methods is a difficult problem.

\subsection{First Experiment}


\global\long\def\strategyCDnet{\ensuremath{\textrm{CDnet}}}
\global\long\def\strategyCDnetCategory{\ensuremath{\textrm{CDnet}^*}}
\global\long\def\strategyMeanPerformance{\ensuremath{\textrm{mean-P}}}
\global\long\def\strategyMeanPerformanceCategory{\ensuremath{\textrm{mean-P}^*}}
\global\long\def\strategySemanticPerformance{\ensuremath{\textrm{sem-P}^\dagger}}
\global\long\def\strategySemanticPerformanceCategory{\ensuremath{\textrm{sem-P}^{\dagger*}}}
\global\long\def\strategySemanticDistance{\ensuremath{\textrm{sem-d}^\dagger}}
\global\long\def\strategySemanticDistanceCategory{\ensuremath{\textrm{sem-d}^{\dagger*}}}
\global\long\def\strategyAverage{\ensuremath{\textrm{avg}^\dagger}}
\global\long\def\strategyAverageCategory{\ensuremath{\textrm{avg}^{\dagger*}}}
\global\long\def\strategyAll{\ensuremath{\textrm{all}^\dagger}}

\newcommand{\strategyDefinition}[1]{
    \textcolor{RoyalBlue}{[#1]}\xspace
}

The aim of the \first experiment is to compare a variety of $11$ very intuitive strategies. 
\strategyDefinition{\strategyCDnet} is the ranking given on \CDnetPlatform\footnote{
    This is our sole strategy that is not leave-one-video-out, as all our input videos are involved in this ranking. The related results could be optimistic.
} at an arbitrarily chosen moment\footnote{
    Version of March, 11th 2025. This remark is needed because the overall ranking of \CDnetPlatform does not satisfy the \first axiom of \cite{Pierard2025Foundations}: it is not stable. The relative order between the \numMethodsBGS methods, can be influenced by the presence of other methods (\eg, the supervised ones) on the platform, and could change if a method was added or removed from the platform.
}. It is not sensitive to application-specific preferences. 
\strategyDefinition{\strategyMeanPerformance} is the ranking based on the summarization~\cite{Pierard2020Summarizing} of the performances determined on the 52 other videos. 
\strategyDefinition{\strategySemanticPerformance} is the ranking based on the performance predicted from, on the one hand, the distribution of semantic labels in the input video and, on the other hand, the probabilities of false/true background/foreground conditionally to the semantic label for the other videos.  
\strategyDefinition{\strategySemanticDistance} is the ranking known for the closest video in terms of semantic characterization (Bhattacharyya distance between distributions of semantic labels).
\strategyDefinition{\strategyAverage} is the ranking based on the summarization~\cite{Pierard2020Summarizing} of the performances predicted for \strategyCDnet\footnote{
    Because the summarization only accepts performances as input, and not real values, we need a solution to convert a value into a performance. For any value $v\in[l,u]$, we assign the performance $\aPerformance$ such that $\aPerformance(\{\sampleTN\})=\aPerformance(\{\sampleTP\})=\frac{1}{2}\frac{v-l}{u-l}$ and $\aPerformance(\{\sampleFP\})=\aPerformance(\{\sampleFN\})=\frac{1}{2}\frac{u-v}{u-l}$, as $\rankingScore(\aPerformance)$ is strictly monotonously increasing with $v$ for all ranking scores $\rankingScore$.
}, \strategyMeanPerformance, \strategySemanticPerformance, and \strategySemanticDistance. 
\strategyDefinition{$\ldots^*$}: 
each of these five strategies comes with a variant that is category-specific and that is denoted by $^*$.
\strategyDefinition{\strategyAll} is the ranking based on the summarization~\cite{Pierard2020Summarizing} of the performances predicted for \strategyAverage ~and \strategyAverageCategory. 

\begin{figure}
\center 
\subfloat[Hybrid strategy maximizing $mean(\tau)$.]{
    \includegraphics[scale=0.24]{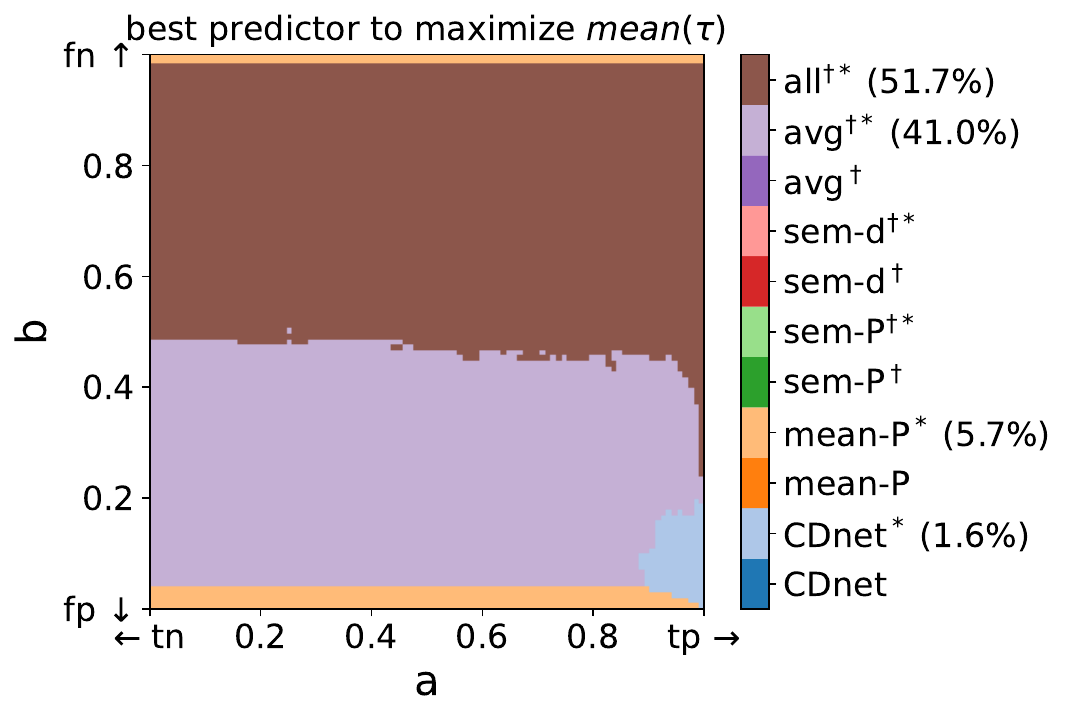}
    \includegraphics[scale=0.24]{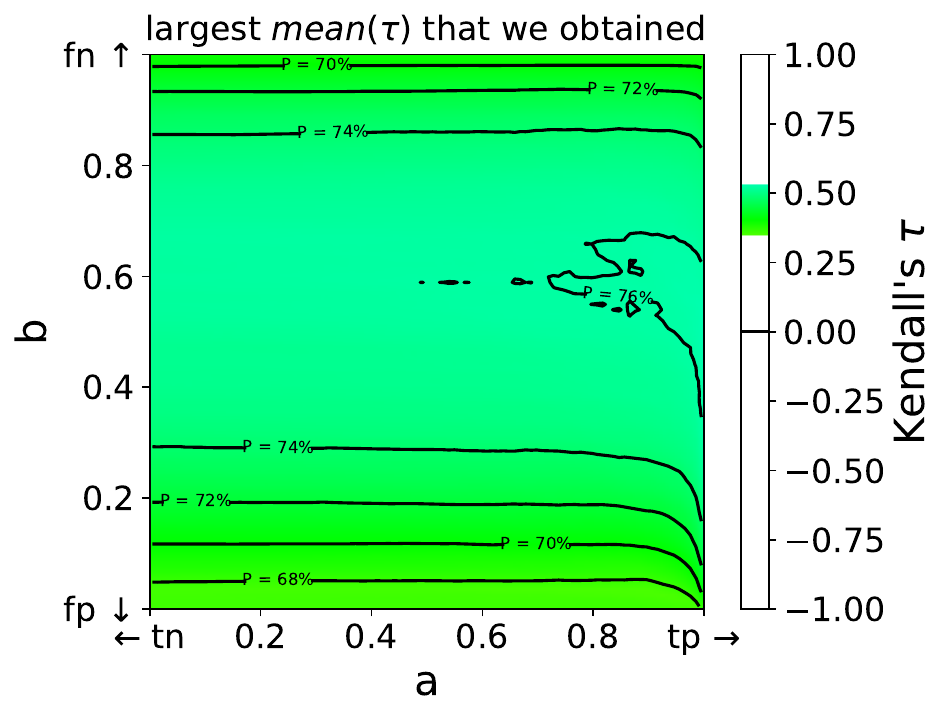}

}\\
\subfloat[Hybrid strategy maximizing $min(\tau)$.]{
    \includegraphics[scale=0.24]{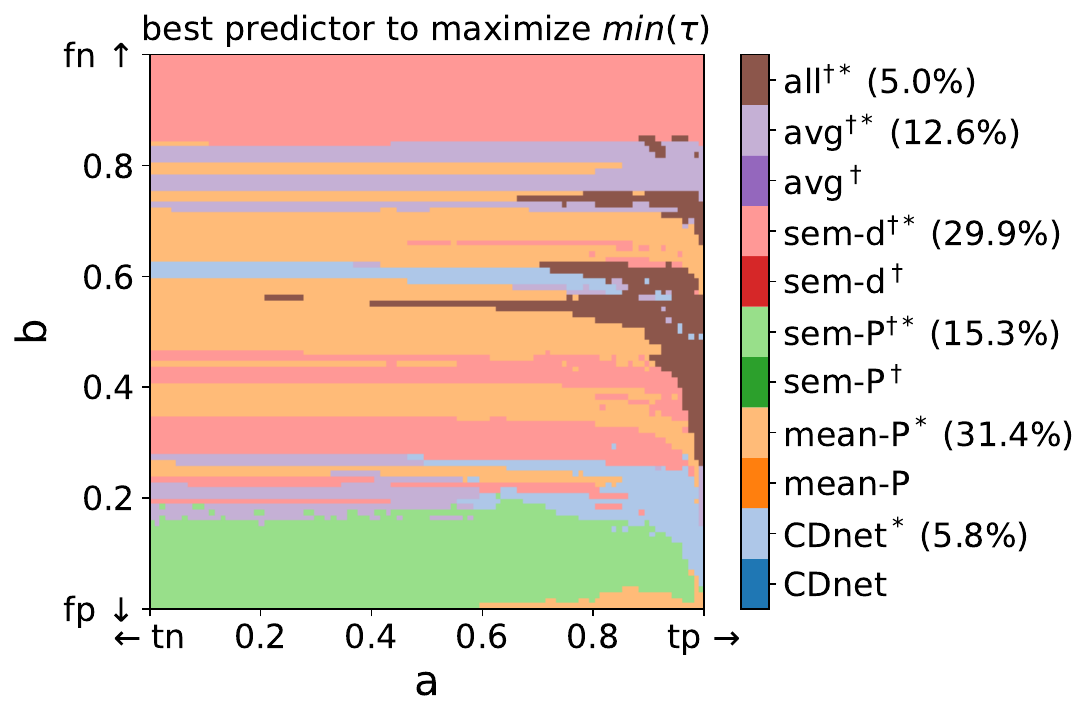}
    \includegraphics[scale=0.24]{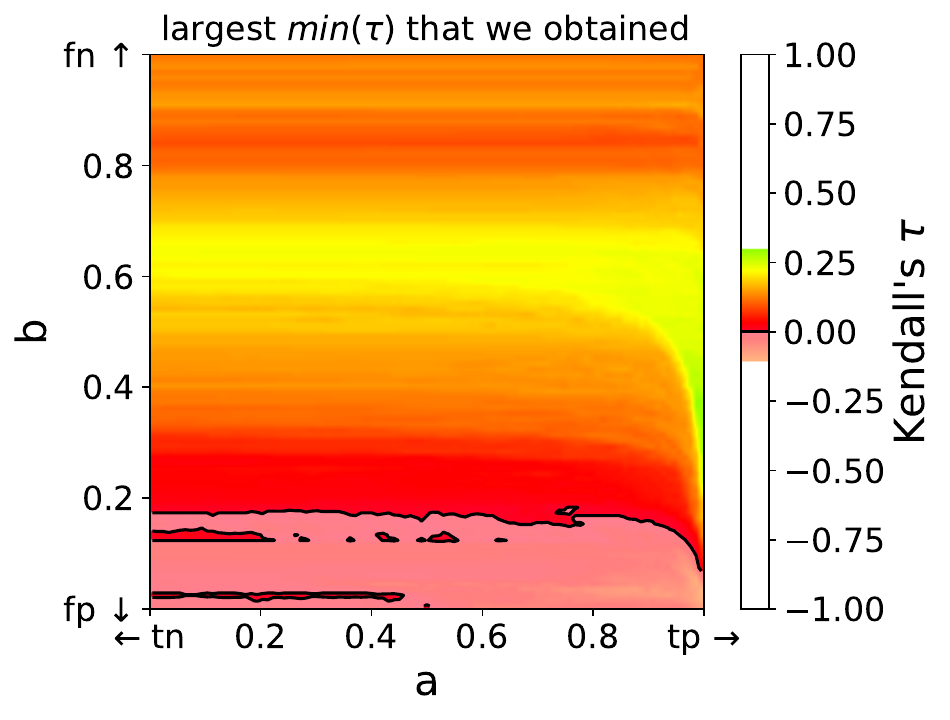}

}
\caption{Results of our \first experiment.\label{fig:results-expe-1}}

\end{figure}

The results are shown in \cref{fig:results-expe-1}. We see that, when we want to maximize $mean(\tau)$, it is clearly useful to consider hybrid strategies, choosing one of the basic strategies according to the application-specific preferences: none of the $11$ strategies tested in this \first experiment is the best in more than $51.7\%$ of the \tile. Moreover, we see that for the vast part of the \tile, it is best to opt for strategies based on the summarization of performances predicted in various ways. The maximum $mean(\tau)$ obtained is relatively independent of application-specific preferences (the \tile is all in green shades). When we want to maximize $min(\tau)$, the strategies to select are, over the majority of the \tile, strategies that exploit knowledge of the category to which the input video belongs.

\subsection{Second Experiment}


\global\long\def\strategyMeanRankCategory{\ensuremath{\textrm{mean-R}^*}}

Now, we aim to establish something simpler, and easier to interpret, than the first experiment. We study the vertical behavior of the \tile through a parametric family of strategies. We focus on strategies that exploit the knowledge of the category, but put semantics aside. This experiment involves $11$ strategies to predict rankings. 
\strategyDefinition{$\strategyMeanRankCategory(\frac12,x)$} is the category-specific weighted\footnote{
    We weight videos as in \CDnetPlatform: all categories have the same weight, and all videos in a given category have also the same weight. The video on which the rankings are predicted has a zero weight.
} arithmetic mean of the ranks induced by the canonical ranking scores $\canonicalRankingScore$ with $a=\frac12$ and $b=x$. 

\begin{figure}
\center 
\subfloat[Hybrid strategy maximizing $mean(\tau)$.]{
    \includegraphics[scale=0.24]{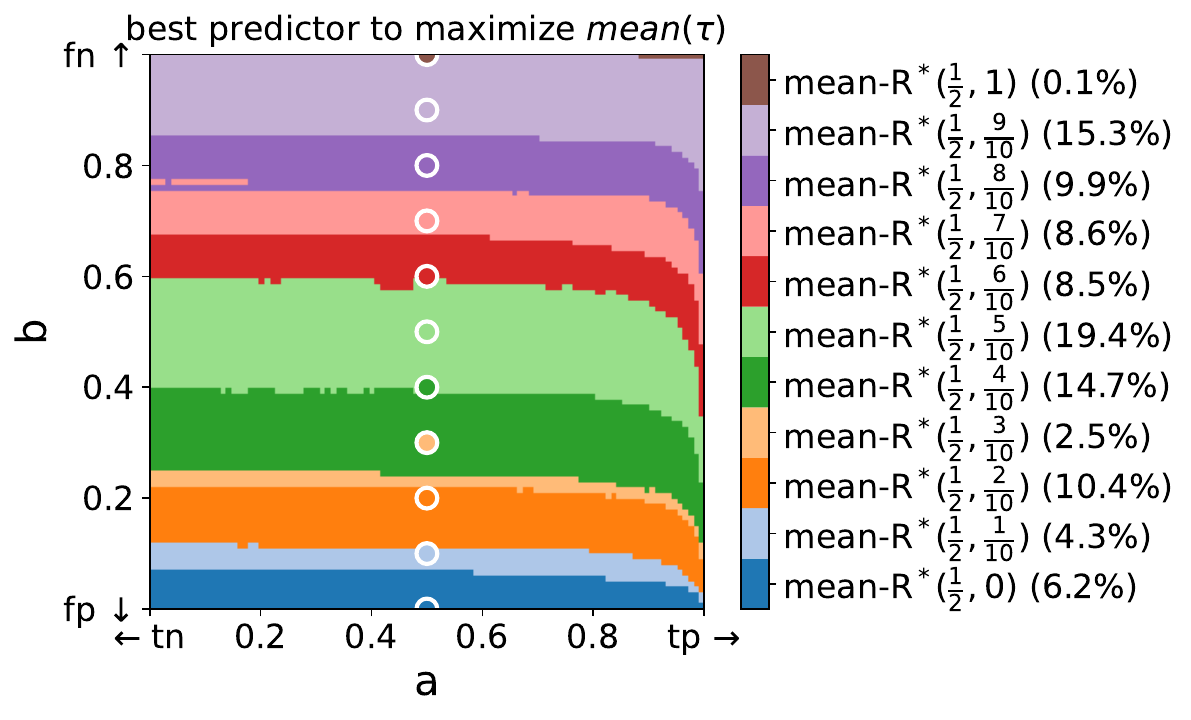}
    \includegraphics[scale=0.24]{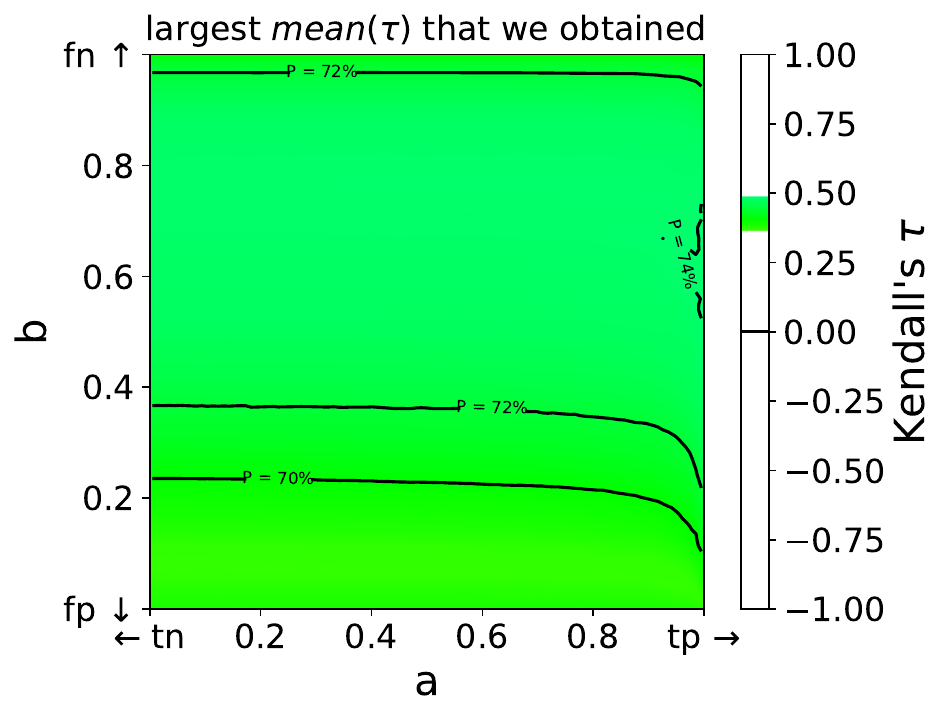}

}\\
\subfloat[Hybrid strategy maximizing $min(\tau)$.]{
    \includegraphics[scale=0.24]{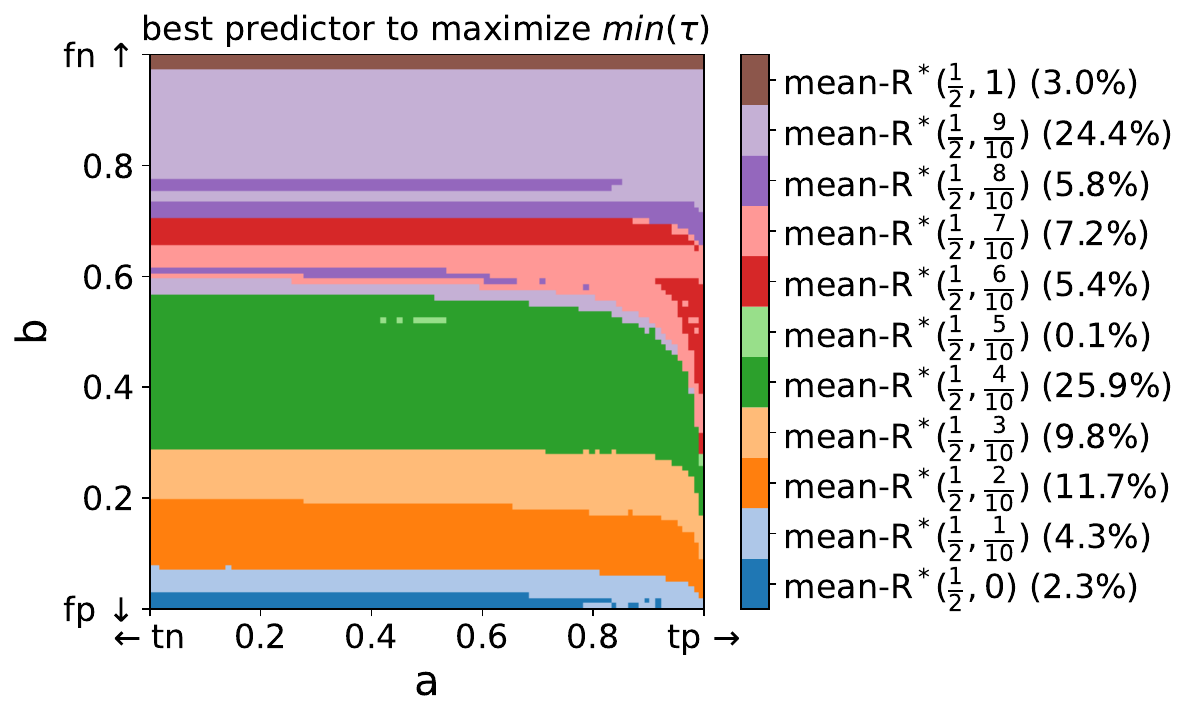}
    \includegraphics[scale=0.24]{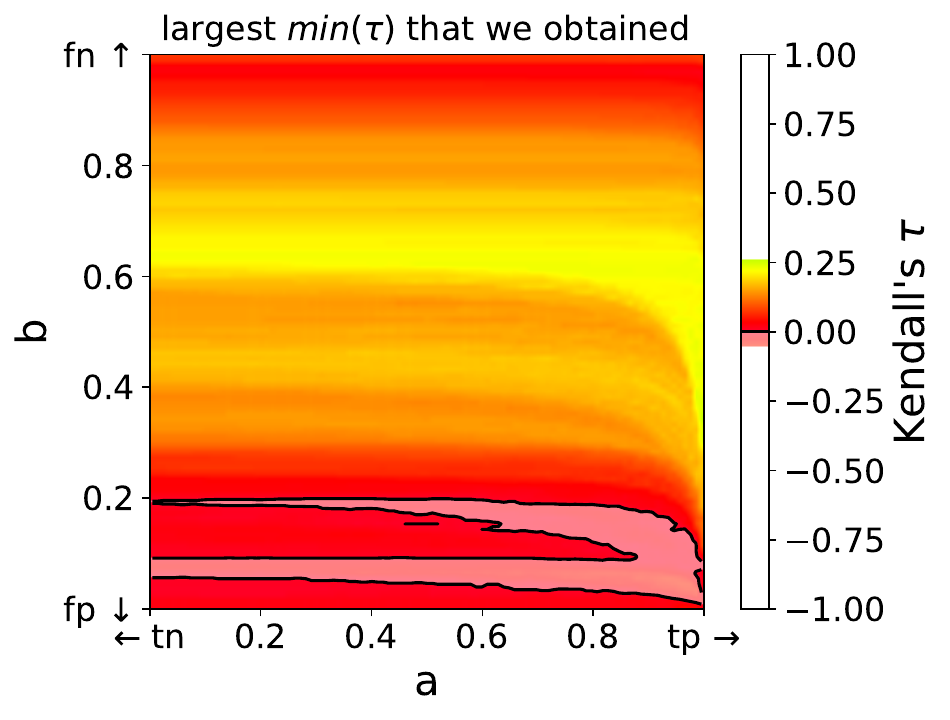}

}
\caption{Results of our \second experiment.\label{fig:results-expe-2}}

\end{figure}

Results are shown in \cref{fig:results-expe-2} for the discrete set of values $x\in\{0, \frac{1}{10}, \frac{2}{10} \ldots 1\}$. When we want to maximize $mean(\tau)$, we see that all $11$ strategies are useful. Moreover, the point $(a,b)=(\frac12,x)$ lies in, or is not far from, the area of the \tile where the strategy $\strategyMeanRankCategory(\frac12,x)$ is the best (see the colored points). The average $\tau$ achievable with these strategies is close to the average $\tau$ obtained with the strategies of the first experiment. When we want to maximize $min(\tau)$, our conclusions are similar to those given for the maximization of $mean(\tau)$. This paves the way for an even more refined hybrid strategy, combining not only $11$ strategies as in this experiment, but an infinite number of them (one for each $a$ and $b$), which is done in the \third experiment.

\subsection{Third Experiment}


\global\long\def\strategyMeanPerformance{\ensuremath{\textrm{mean-P}}}
\global\long\def\strategyMeanPerformanceCategory{\ensuremath{\textrm{mean-P}^*}}
\global\long\def\strategyMeanValue{\ensuremath{\textrm{mean-V}}}
\global\long\def\strategyMeanValueCategory{\ensuremath{\textrm{mean-V}^*}}
\global\long\def\strategyMeanRank{\ensuremath{\textrm{mean-R}}}
\global\long\def\strategyMeanRankCategory{\ensuremath{\textrm{mean-R}^*}}
\global\long\def\strategyMedianValue{\ensuremath{\textrm{med-V}}}
\global\long\def\strategyMedianValueCategory{\ensuremath{\textrm{med-V}^*}}
\global\long\def\strategyMedianRank{\ensuremath{\textrm{med-R}}}
\global\long\def\strategyMedianRankCategory{\ensuremath{\textrm{med-R}^*}}

Now, we aim to explore two questions: (1)~is it better to compute average performances, average values, or average ranks? 
(2)~is it more suitable to compute means or medians? 
This experiment involves $10$ strategies. 
\strategyDefinition{\strategyMeanPerformance} is the ranking based on the summarization~\cite{Pierard2020Summarizing} of the performances determined on the 52 other videos. 
\strategyDefinition{\strategyMeanValue} is the weighted arithmetic mean of the values taken by the canonical ranking score, corresponding to the application-specific preferences. 
\strategyDefinition{\strategyMeanRank} is the weighted arithmetic mean of the ranks induced by the canonical ranking score, corresponding to the application-specific preferences. 
\strategyDefinition{\strategyMedianValue} is similar to \strategyMeanValue, except that we take the median instead of the mean. 
\strategyDefinition{\strategyMedianRank} is similar to \strategyMeanRank, except that we take the median instead of the mean. 
\strategyDefinition{$\ldots^*$}: 
as in the \first experiment, each of these five strategies comes with a variant that is category-specific and that is denoted by $^*$.

\begin{figure}
\center 
\subfloat[Hybrid strategy maximizing $mean(\tau)$.]{
    \includegraphics[scale=0.24]{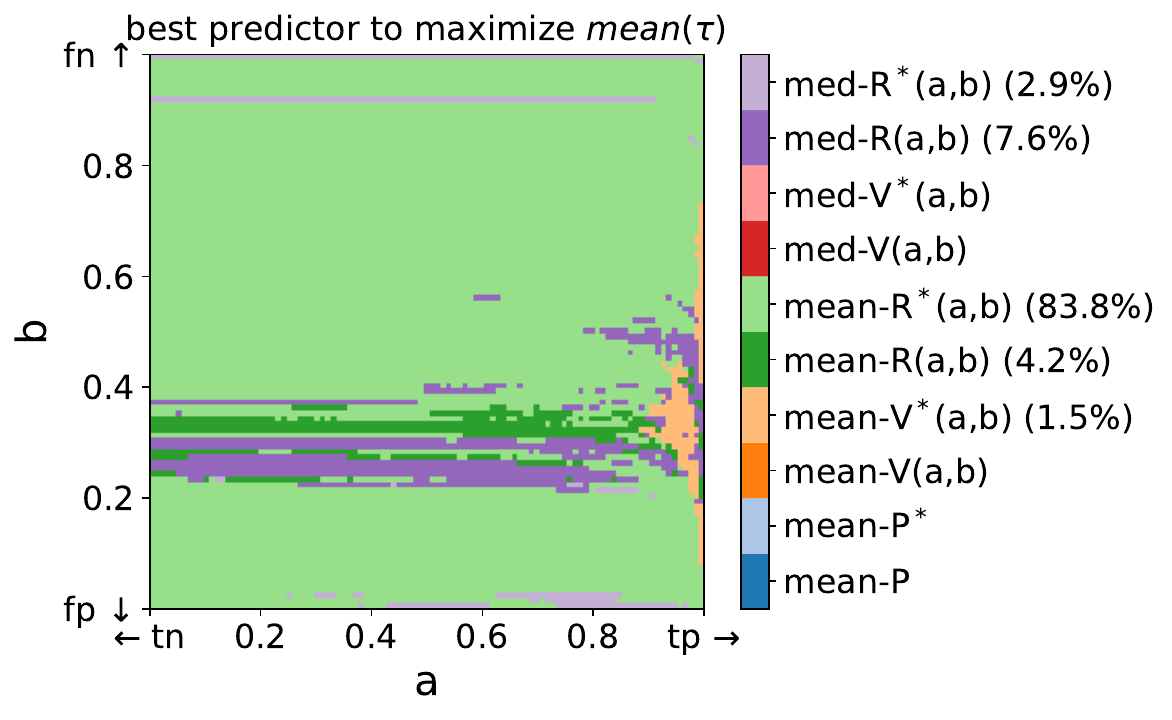}
    \includegraphics[scale=0.24]{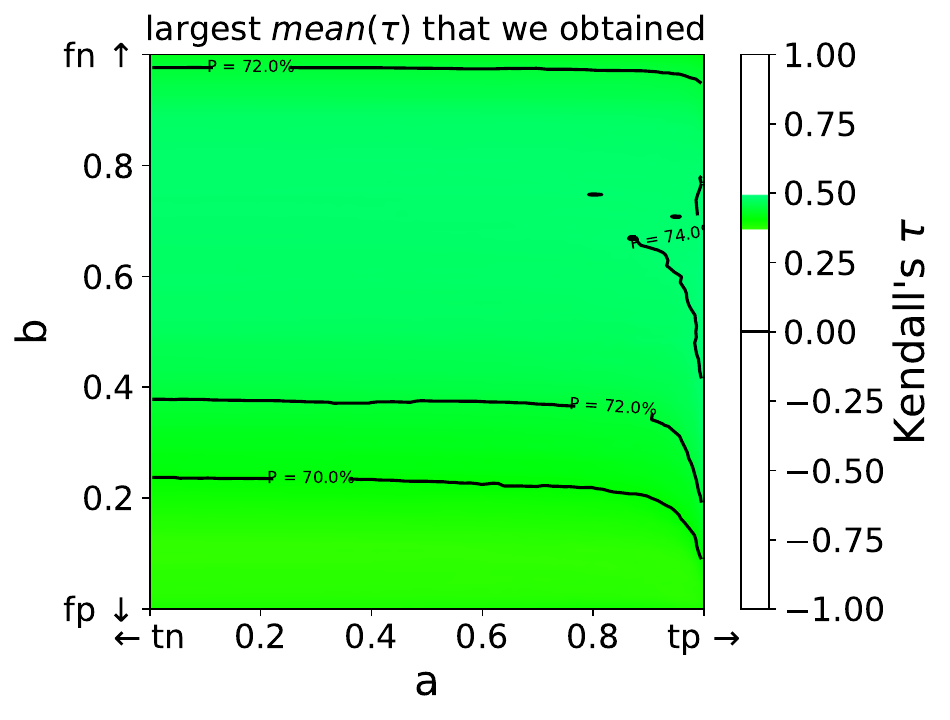}

}\\
\subfloat[Hybrid strategy maximizing $min(\tau)$.]{
    \includegraphics[scale=0.24]{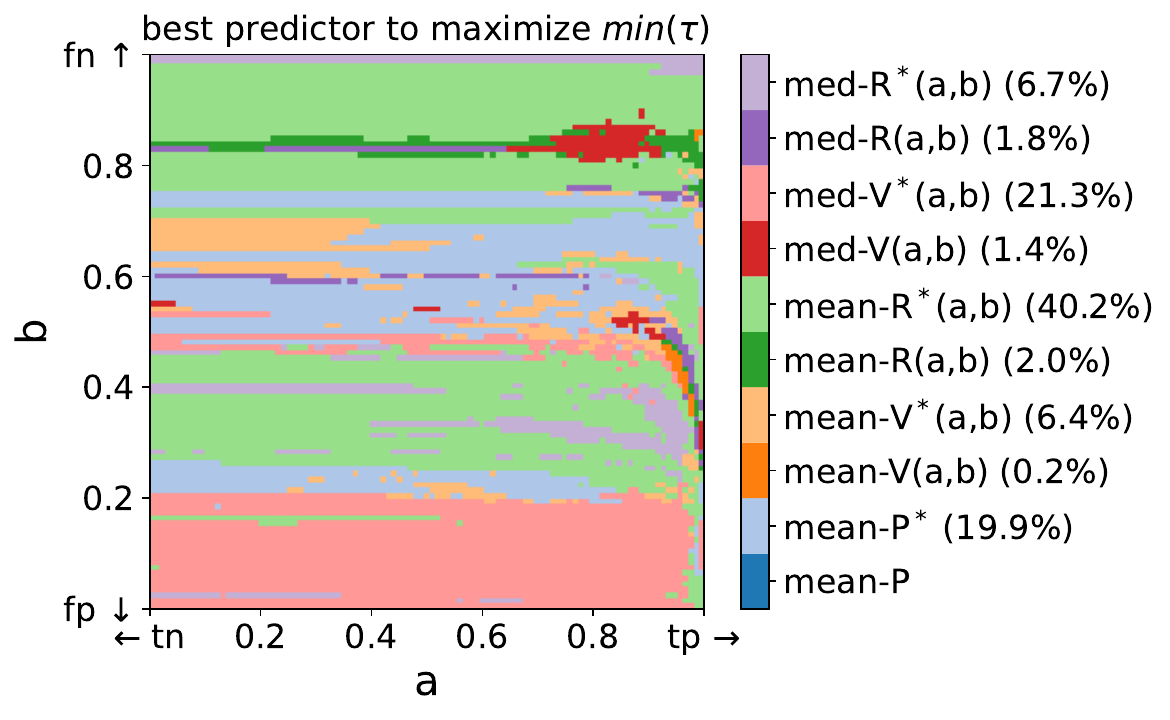}
    \includegraphics[scale=0.24]{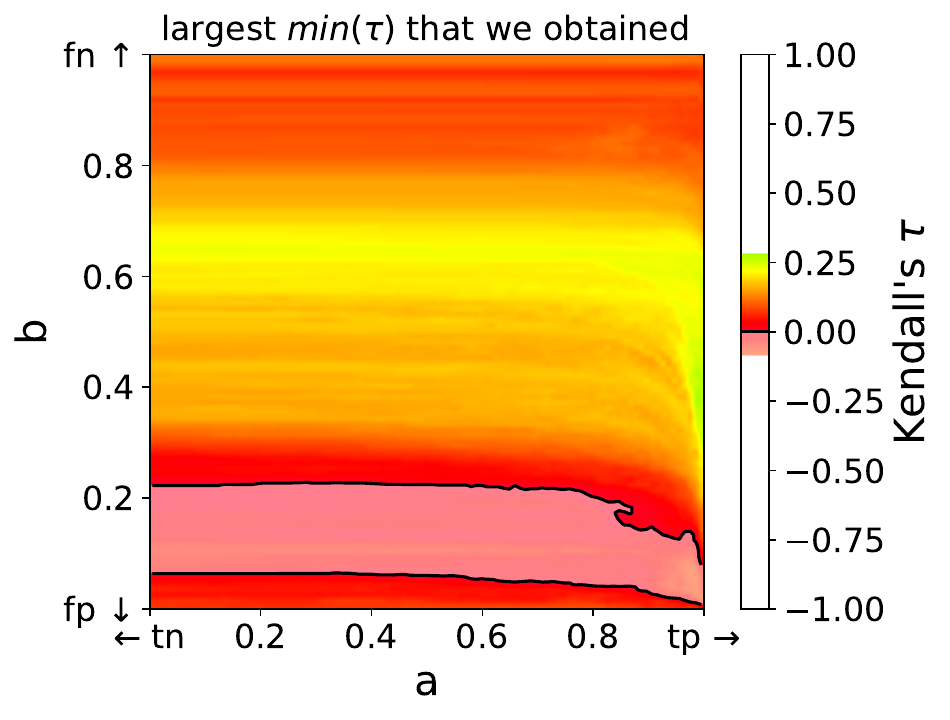}

}
\caption{Results of our \third experiment.\label{fig:results-expe-3}}

\end{figure}

The results of the \third experiment are displayed 
in \cref{fig:results-expe-3}. When we want to maximize $mean(\tau)$, in $83.8\%$ of the \tile, \strategyMeanRankCategory~is the best strategy among the $10$. 
When we want to maximize $min(\tau)$, for most of the \tile, it is better to take only strategies that exploit videos within the same category. The $mean(\tau)$ and $min(\tau)$ \tiles are 
very similar to the ones we obtained in our first two experiments.

\section{Conclusion}

This work introduces a methodology for comparing strategies for the prediction of rankings of computer vision methods (\eg, \bgs methods) on new domains (\eg, videos), \wrt application-specific preferences, by exploiting the knowledge that can be found in publicly available multi-domain rankings (\eg, \CDnetPlatform). The methodology, as presented for a problem similar to the two-class classification, takes advantage of a recently introduced visualization tool called ``Tile''~\cite{Pierard2024TheTile-arxiv,Halin2024AHitchhikers-arxiv}, which is based on a theoretical framework for performances~\cite{Pierard2025Foundations}. In the case that we studied, we have shown that the performance-based rankings of computer vision methods on new domains is far from being an easy or solved problem. Because of its practical importance, we hope that this work will stimulate the community to investigate and develop more powerful strategies, and that our proposed methodology will help to establish their effectiveness.

\section*{Acknowledgment}

S. Pi{\'e}rard, A. Deli{\`e}ge, and A. Halin are funded respectively by (1) the Service Public de Wallonie (SPW) Recherche (grant 8573, ReconnAIssance project), (2) ULi{\`e}ge (project DESTINA), and (3) the SPW EER, Wallonia, Belgium (grant 2010235, ARIAC by \href{https://www.digitalwallonia.be/en/}{DIGITALWALLONIA4.AI}).

\newpage

\appendix 

\subsection{Useful resources}

Useful resources can be found at:
\begin{center}
\url{https://github.com/pierard/performance}
\end{center}

\subsection{Enlarged version of~\cref{fig:experiment_introductory_example_blizzard_S}}
 
\begin{figure*}[t!]
\global\long\def\strategyCDnet{\ensuremath{\mathrm{CDnet}}}
\global\long\def\strategySemanticDistance{\ensuremath{\mathrm{sem-d}^\dagger}}
\global\long\def\strategySemanticDistanceCategory{\ensuremath{\mathrm{sem-d}^{\dagger*}}}
\begin{centering}
    \hfill{}
    \subfloat[Strategy \strategyCDnet]{
        \begin{centering}
        \includegraphics[width=0.4\columnwidth]{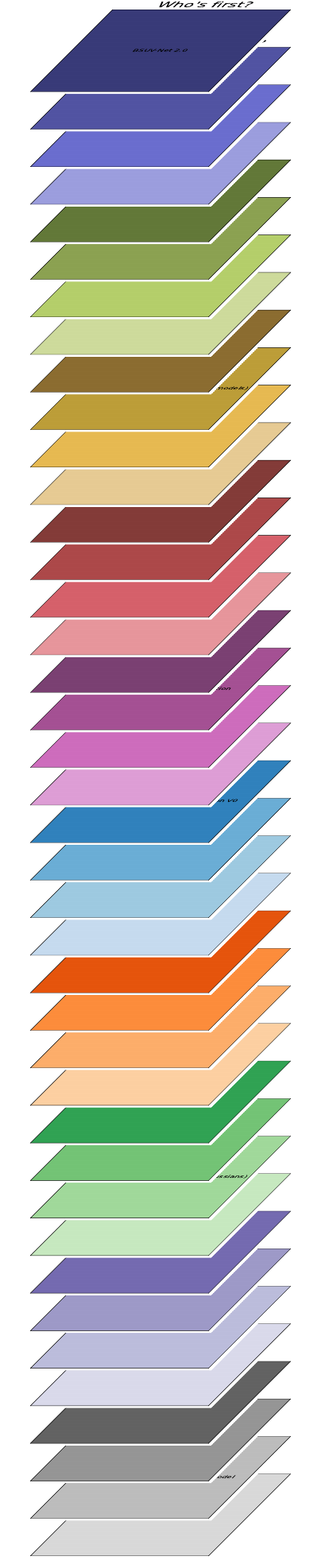}
        \par\end{centering}
    }
    \hfill{}\hfill{}
    \subfloat[Strategy \strategySemanticDistance]{
        \begin{centering}
        \includegraphics[width=0.4\columnwidth]{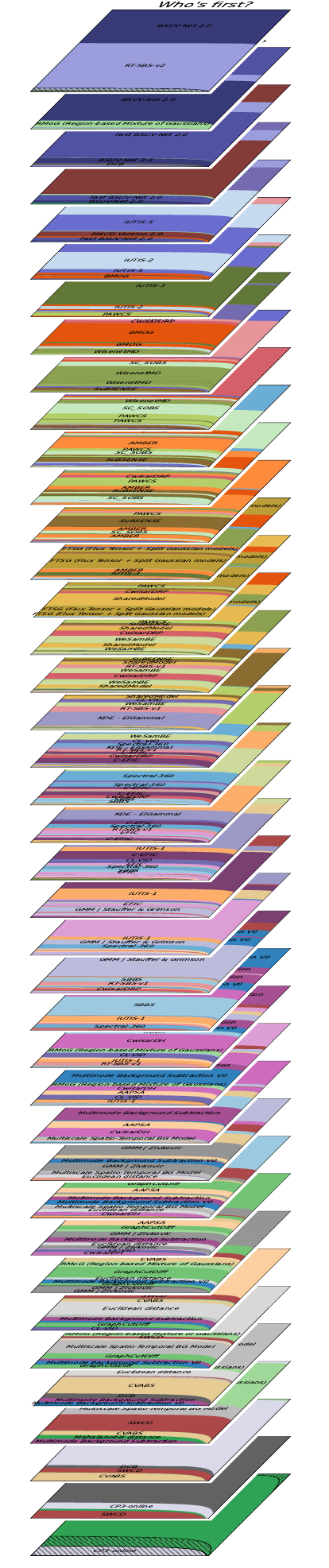}
        \par\end{centering}
    }
    \hfill{}\hfill{}
    \subfloat[Strategy \strategySemanticDistanceCategory]{
        \begin{centering}
        \includegraphics[width=0.4\columnwidth]{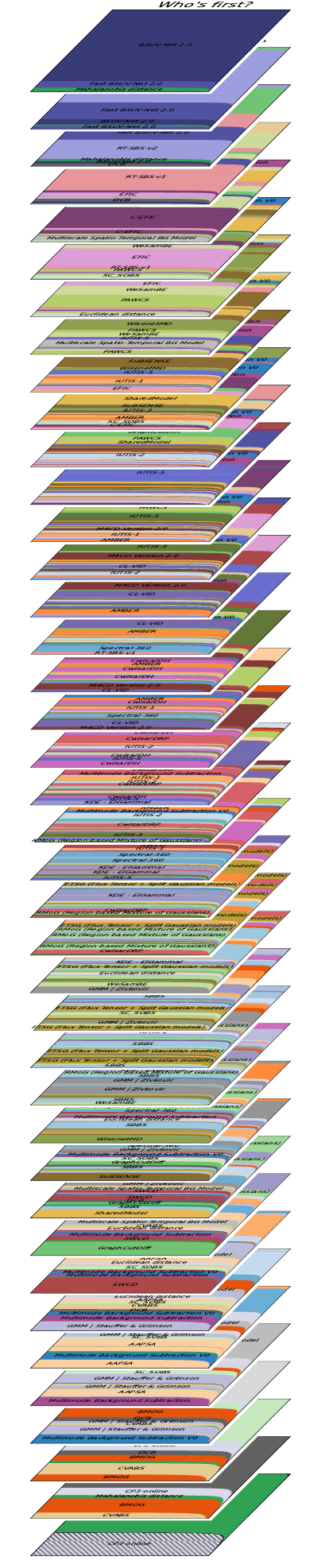}
        \par\end{centering}
    }
    \hfill{}\hfill{}
    \subfloat[Real rankings]{
        \begin{centering}
        \includegraphics[width=0.4\columnwidth]{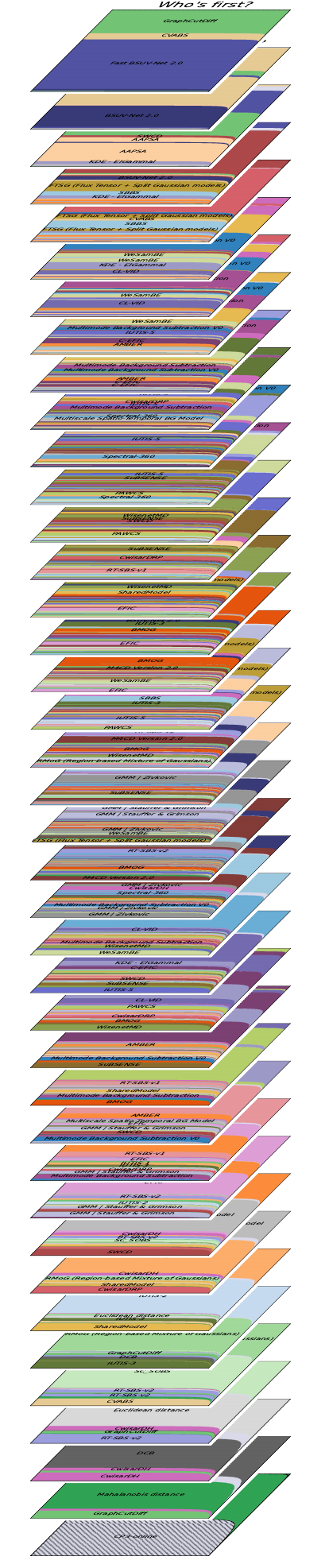}
        \par\end{centering}
    }
    \hfill{}
\par\end{centering}
\begin{centering}
    \hfill{}
    \subfloat[Strategy \strategyCDnet]{
        \begin{centering}
        \begin{minipage}[t]{0.39\columnwidth}%
            \begin{center}
            \includegraphics[scale=0.27]{images/experiment_introductory_example_blizzard/correlation_tile_strategy_1a}
            \par\end{center}%
        \end{minipage}
        \par\end{centering}
    }
    \hfill{}\hfill{}
    \subfloat[Strategy \strategySemanticDistance]{
        \begin{centering}
        \begin{minipage}[t]{0.39\columnwidth}%
            \begin{center}
            \includegraphics[scale=0.27]{images/experiment_introductory_example_blizzard/correlation_tile_strategy_7a}
            \par\end{center}%
        \end{minipage}
        \par\end{centering}
    }
    \hfill{}\hfill{}
    \subfloat[Strategy \strategySemanticDistanceCategory]{
        \begin{centering}
        \begin{minipage}[t]{0.39\columnwidth}%
            \begin{center}
            \includegraphics[scale=0.27]{images/experiment_introductory_example_blizzard/correlation_tile_strategy_8a}
            \par\end{center}%
        \end{minipage}
        \par\end{centering}
    }
    \hfill{}\hfill{}
    \subfloat[Choice \wrt $\randVarImportance$.]{
        \begin{centering}
        \begin{minipage}[t]{0.39\columnwidth}%
            \begin{center}
            \includegraphics[scale=0.27]{images/experiment_introductory_example_blizzard/best_strategy}
            \par\end{center}%
        \end{minipage}
        \par\end{centering}
    }
    \hfill{}
\par\end{centering}
\caption{
    Enlarged version of \cref{fig:experiment_introductory_example_blizzard_S}.
    \label{fig:experiment_introductory_example_blizzard_XL}
}
\end{figure*}

\cref{fig:experiment_introductory_example_blizzard_XL} is an enlarged version of \cref{fig:experiment_introductory_example_blizzard_S} on which it is easier to observe the $\numMethodsBGS$ layers of the mille-feuilles. 

\subsection{\bgs methods considered in \cref{sec:application}}

We considered $40$ unsupervised \bgs methods. More precisely, we selected those for which the output masks were available on \CDnetPlatform, for all videos of \CDnetMMXIV, on March 11th 2025, and excluded those tagged as ``supervised'' on the platform.
~\\

{
\scriptsize
    \begin{enumerate}
        \item \verb@BSUV-Net 2.0@
        \item \verb@Fast BSUV-Net 2.0@
        \item \verb@IUTIS-5@
        \item \verb@RT-SBS-v2@
        \item \verb@IUTIS-3@
        \item \verb@WisenetMD@
        \item \verb@PAWCS@
        \item \verb@WeSamBE@
        \item \verb@SuBSENSE@
        \item \verb@FTSG (Flux Tensor with Split Gaussian mdoels)@
        \item \verb@SharedModel@
        \item \verb@CVABS@
        \item \verb@M4CD Version 2.0@
        \item \verb@SWCD@
        \item \verb@CwisarDRP@
        \item \verb@RT-SBS-v1@
        \item \verb@C-EFIC@
        \item \verb@Multimode Background Subtraction@
        \item \verb@CwisarDH@
        \item \verb@EFIC@
        \item \verb@Multimode Background Subtraction Version 0 (MBS V0)@
        \item \verb@Spectral-360@
        \item \verb@Sample based background subtractor (SBBS)@
        \item \verb@IUTIS-2@
        \item \verb@BMOG@
        \item \verb@AMBER@
        \item \verb@IUTIS-1@
        \item \verb@AAPSA@
        \item \verb@Mahalanobis distance@
        \item \verb@GraphCutDiff@
        \item \verb@RMoG (Region-based Mixture of Gaussians)@
        \item \verb@SC_SOBS@
        \item \verb@CL-VID@
        \item \verb@KDE - ElGammal@
        \item \verb@GMM | Stauffer & Grimson@
        \item \verb@CP3-online@
        \item \verb@DCB@
        \item \verb@GMM | Zivkovic@
        \item \verb@Multiscale Spatio-Temporal BG Model@
        \item \verb@Euclidean distance@
    \end{enumerate}
}

\subsection{Detailed results for the first experiment}

For the 11 strategies involved in our \first experiment, we provide the correlation \tiles  for each domain (the \numVideos videos of \CDnetMMXIV), as well as the average (mean correlation \tile) and worst-case (minimum correlation \tile) over all domains. These are the intermediate results that are summarized in \cref{fig:results-expe-1}.

\begin{itemize}
    \item \cref{fig:strategyCDnet}: detailed results for \strategyCDnet
    \item \cref{fig:strategyCDnetCategory}: detailed results for \strategyCDnetCategory
    \item \cref{fig:strategyMeanPerformance}: detailed results for \strategyMeanPerformance
    \item \cref{fig:strategyMeanPerformanceCategory}: detailed results for \strategyMeanPerformanceCategory
    \item \cref{fig:strategySemanticPerformance}: detailed results for \strategySemanticPerformance
    \item \cref{fig:strategySemanticPerformanceCategory}: detailed results for \strategySemanticPerformanceCategory
    \item \cref{fig:strategySemanticDistance}: detailed results for \strategySemanticDistance
    \item \cref{fig:strategySemanticDistanceCategory}: detailed results for \strategySemanticDistanceCategory
    \item \cref{fig:strategyAverage}: detailed results for \strategyAverage
    \item \cref{fig:strategyAverageCategory}: detailed results for \strategyAverageCategory
    \item \cref{fig:strategyAll}: detailed results for \strategyAll
\end{itemize}

\newcommand{\plate}[1]{%
    \setlength{\unitlength}{1mm}
    \begin{picture}(300,350)(0,0)
    
    \multiput(5,225)(40,0){2}{\line(0,1){125}}%
    \multiput(5,225)(0,125){2}{\line(1,0){40}}%
    \put(0,225){\rotatebox{90.0}{BAD WEATHER}}
    \put(10,230){\includegraphics[width=30mm]{#1/badWeather_blizzard.pdf}}
    \put(10,260){\includegraphics[width=30mm]{#1/badWeather_skating.pdf}}
    \put(10,290){\includegraphics[width=30mm]{#1/badWeather_snowFall.pdf}}
    \put(10,320){\includegraphics[width=30mm]{#1/badWeather_wetSnow.pdf}}
    
    \multiput(55,225)(40,0){2}{\line(0,1){125}}%
    \multiput(55,225)(0,125){2}{\line(1,0){40}}%
    \put(50,225){\rotatebox{90.0}{BASELINE}}
    \put(60,230){\includegraphics[width=30mm]{#1/baseline_PETS2006.pdf}}
    \put(60,260){\includegraphics[width=30mm]{#1/baseline_highway.pdf}}
    \put(60,290){\includegraphics[width=30mm]{#1/baseline_office.pdf}}
    \put(60,320){\includegraphics[width=30mm]{#1/baseline_pedestrians.pdf}}
    
    \multiput(105,225)(40,0){2}{\line(0,1){125}}%
    \multiput(105,225)(0,125){2}{\line(1,0){40}}%
    \put(100,225){\rotatebox{90.0}{LOW FRAMERATE}}
    \put(110,230){\includegraphics[width=30mm]{#1/lowFramerate_port_0_17fps.pdf}}
    \put(110,260){\includegraphics[width=30mm]{#1/lowFramerate_tramCrossroad_1fps.pdf}}
    \put(110,290){\includegraphics[width=30mm]{#1/lowFramerate_tunnelExit_0_35fps.pdf}}
    \put(110,320){\includegraphics[width=30mm]{#1/lowFramerate_turnpike_0_5fps.pdf}}
    
    \multiput(155,225)(40,0){2}{\line(0,1){125}}%
    \multiput(155,225)(0,125){2}{\line(1,0){40}}%
    \put(150,225){\rotatebox{90.0}{TURBULENCE}}
    \put(160,230){\includegraphics[width=30mm]{#1/turbulence_turbulence0.pdf}}
    \put(160,260){\includegraphics[width=30mm]{#1/turbulence_turbulence1.pdf}}
    \put(160,290){\includegraphics[width=30mm]{#1/turbulence_turbulence2.pdf}}
    \put(160,320){\includegraphics[width=30mm]{#1/turbulence_turbulence3.pdf}}
    
    \multiput(205,225)(40,0){2}{\line(0,1){125}}%
    \multiput(205,225)(0,125){2}{\line(1,0){40}}%
    \put(200,225){\rotatebox{90.0}{CAMERA JITTER}}
    \put(210,230){\includegraphics[width=30mm]{#1/cameraJitter_badminton.pdf}}
    \put(210,260){\includegraphics[width=30mm]{#1/cameraJitter_boulevard.pdf}}
    \put(210,290){\includegraphics[width=30mm]{#1/cameraJitter_sidewalk.pdf}}
    \put(210,320){\includegraphics[width=30mm]{#1/cameraJitter_traffic.pdf}}

    \multiput(255,225)(40,0){2}{\line(0,1){125}}%
    \multiput(255,225)(0,125){2}{\line(1,0){40}}%
    \put(250,225){\rotatebox{90.0}{PTZ}}
    \put(260,230){\includegraphics[width=30mm]{#1/PTZ_continuousPan.pdf}}
    \put(260,260){\includegraphics[width=30mm]{#1/PTZ_intermittentPan.pdf}}
    \put(260,290){\includegraphics[width=30mm]{#1/PTZ_twoPositionPTZCam.pdf}}
    \put(260,320){\includegraphics[width=30mm]{#1/PTZ_zoomInZoomOut.pdf}}
    
    \multiput(5,180)(215,0){2}{\line(0,1){35}}%
    \multiput(5,180)(0,35){2}{\line(1,0){215}}%
    \put(5,218){INTERMITTENT OBJECT MOTION}
    \put(10,185){\includegraphics[width=30mm]{#1/intermittentObjectMotion_abandonedBox.pdf}}
    \put(45,185){\includegraphics[width=30mm]{#1/intermittentObjectMotion_parking.pdf}}
    \put(80,185){\includegraphics[width=30mm]{#1/intermittentObjectMotion_sofa.pdf}}
    \put(115,185){\includegraphics[width=30mm]{#1/intermittentObjectMotion_streetLight.pdf}}
    \put(150,185){\includegraphics[width=30mm]{#1/intermittentObjectMotion_tramstop.pdf}}
    \put(185,185){\includegraphics[width=30mm]{#1/intermittentObjectMotion_winterDriveway.pdf}}
    
    \multiput(5,135)(215,0){2}{\line(0,1){35}}%
    \multiput(5,135)(0,35){2}{\line(1,0){215}}%
    \put(5,173){DYNAMIC BACKGROUND}
    \put(10,140){\includegraphics[width=30mm]{#1/dynamicBackground_boats.pdf}}
    \put(45,140){\includegraphics[width=30mm]{#1/dynamicBackground_canoe.pdf}}
    \put(80,140){\includegraphics[width=30mm]{#1/dynamicBackground_fall.pdf}}
    \put(115,140){\includegraphics[width=30mm]{#1/dynamicBackground_fountain01.pdf}}
    \put(150,140){\includegraphics[width=30mm]{#1/dynamicBackground_fountain02.pdf}}
    \put(185,140){\includegraphics[width=30mm]{#1/dynamicBackground_overpass.pdf}}
    
    \multiput(5,90)(215,0){2}{\line(0,1){35}}%
    \multiput(5,90)(0,35){2}{\line(1,0){215}}%
    \put(5,128){SHADOW}
    \put(10,95){\includegraphics[width=30mm]{#1/shadow_backdoor.pdf}}
    \put(45,95){\includegraphics[width=30mm]{#1/shadow_bungalows.pdf}}
    \put(80,95){\includegraphics[width=30mm]{#1/shadow_busStation.pdf}}
    \put(115,95){\includegraphics[width=30mm]{#1/shadow_copyMachine.pdf}}
    \put(150,95){\includegraphics[width=30mm]{#1/shadow_cubicle.pdf}}
    \put(185,95){\includegraphics[width=30mm]{#1/shadow_peopleInShade.pdf}}
    
    \multiput(5,45)(215,0){2}{\line(0,1){35}}%
    \multiput(5,45)(0,35){2}{\line(1,0){215}}%
    \put(5,83){NIGHT VIDEOS}
    \put(10,50){\includegraphics[width=30mm]{#1/nightVideos_bridgeEntry.pdf}}
    \put(45,50){\includegraphics[width=30mm]{#1/nightVideos_busyBoulvard.pdf}}
    \put(80,50){\includegraphics[width=30mm]{#1/nightVideos_fluidHighway.pdf}}
    \put(115,50){\includegraphics[width=30mm]{#1/nightVideos_streetCornerAtNight.pdf}}
    \put(150,50){\includegraphics[width=30mm]{#1/nightVideos_tramStation.pdf}}
    \put(185,50){\includegraphics[width=30mm]{#1/nightVideos_winterStreet.pdf}}
    
    \multiput(5,0)(215,0){2}{\line(0,1){35}}%
    \multiput(5,0)(0,35){2}{\line(1,0){215}}%
    \put(5,38){THERMAL}
    \put(25,5){\includegraphics[width=30mm]{#1/thermal_corridor.pdf}}
    \put(60,5){\includegraphics[width=30mm]{#1/thermal_diningRoom.pdf}}
    \put(95,5){\includegraphics[width=30mm]{#1/thermal_lakeSide.pdf}}
    \put(130,5){\includegraphics[width=30mm]{#1/thermal_library.pdf}}
    \put(165,5){\includegraphics[width=30mm]{#1/thermal_park.pdf}}
    
    \put(225,188){WORST CASE: $min(\tau)$}
    \multiput(225,120)(70,0){2}{\line(0,1){65}}%
    \multiput(225,120)(0,65){2}{\line(1,0){70}}%
    \put(230,125){\includegraphics[width=60mm]{#1/global_tile_min_tau.pdf}}
    
    \put(225,113){EXPECTED: $mean(\tau)$}
    \multiput(225,45)(70,0){2}{\line(0,1){65}}%
    \multiput(225,45)(0,65){2}{\line(1,0){70}}%
    \put(230,50){\includegraphics[width=60mm]{#1/global_tile_mean_tau.pdf}}
    
    \put(225,0){\includegraphics[width=70mm]{images/colormap_for_correlations.pdf}}
    
    \end{picture}
}

\newcommand{\figureplate}[3]{%
    \begin{figure*}
        \centering
        \resizebox{\linewidth}{!}{
            \plate{#1}
        }

        \caption{
            Tiles showing the correlations (Kendall's $\tau$), \wrt the application-specific preferences $\randVarImportance$, between the true ranking and the one predicted with the strategy \textcolor{blue}{"#2"}, for \numVideos videos, as well as in the behavior in the worst case and in average. These rankings involve  \numMethodsBGS \bgs methods.
            \label{#3}
        }
    \end{figure*}
}

\figureplate{images/results_predictor_1a}{\strategyCDnet}{fig:strategyCDnet}
\figureplate{images/results_predictor_2a}{\strategyCDnetCategory}{fig:strategyCDnetCategory}
\figureplate{images/results_predictor_3a}{\strategyMeanPerformance}{fig:strategyMeanPerformance}
\figureplate{images/results_predictor_4a}{\strategyMeanPerformanceCategory}{fig:strategyMeanPerformanceCategory}
\figureplate{images/results_predictor_5a}{\strategySemanticPerformance}{fig:strategySemanticPerformance}
\figureplate{images/results_predictor_6a}{\strategySemanticPerformanceCategory}{fig:strategySemanticPerformanceCategory}
\figureplate{images/results_predictor_7a_v3}{\strategySemanticDistance}{fig:strategySemanticDistance}
\figureplate{images/results_predictor_8a_v3}{\strategySemanticDistanceCategory}{fig:strategySemanticDistanceCategory}
\figureplate{images/results_predictor_9a}{\strategyAverage}{fig:strategyAverage}
\figureplate{images/results_predictor_10a}{\strategyAverageCategory}{fig:strategyAverageCategory}
\figureplate{images/results_predictor_11a}{\strategyAll}{fig:strategyAll}

\end{document}